\documentclass{aastex631}
\usepackage[utf8]{inputenc}
\usepackage[T1]{fontenc}
\usepackage{listings}
\usepackage[deluxetable,revtex,bfvec]{math-shortcuts}
\hypersetup{colorlinks, linkcolor=blue, citecolor=blue, urlcolor=blue}
\begin{document}
\title{Characterizing the effects of pulse shape changes on pulsar timing precision}
\author[0000-0003-1082-2342]{Ross J. Jennings}
\altaffiliation{NANOGrav Physics Frontiers Center Postdoctoral Fellow}
\affiliation{Department of Physics and Astronomy, West Virginia University, P.O. Box 6315, Morgantown, WV 26506, USA}
\affiliation{Center for Gravitational Waves and Cosmology, West Virginia University, Chestnut Ridge Research Building, Morgantown, WV 26505, USA}
\author[0000-0002-4049-1882]{James M. Cordes}
\affiliation{Cornell Center for Astrophysics and Planetary Science and Department of Astronomy, Cornell University, Ithaca, NY 14853, USA}
\author[0000-0002-2878-1502]{Shami Chatterjee}
\affiliation{Cornell Center for Astrophysics and Planetary Science and Department of Astronomy, Cornell University, Ithaca, NY 14853, USA}
\date{\today}

\begin{abstract}
Time-of-arrival (TOA) measurements of pulses from pulsars are conventionally made by a template matching algorithm that compares a profile constructed by averaging a finite number of pulses to a long-term average pulse shape. However, the shapes of pulses can and do vary, leading to errors in TOA estimation.  All pulsars show stochastic variations in shape, amplitude, and phase between successive pulses that only partially average out 
in averages of finitely many pulses.  This jitter  phenomenon  will only become more problematic for timing precision as more sensitive telescopes are built. 
We  describe  techniques for  characterizing jitter (and other shape variations) and demonstrate them with data from the Vela pulsar, PSR B0833$-$45. These include partial sum analyses; auto-and cross correlations between templates and profiles and between multifrequency arrival times; and principal component analysis.
We then  quantify how pulse shape changes affect TOA estimates using both analytical and simulation methods on pulse shapes of varying complexity (multiple components). 
These methods can provide the means for improving arrival time precision for many applications, including gravitational wave astronomy using pulsar timing arrays. 
\end{abstract}

\section{Introduction}

Pulsar timing involves making precise measurements of times of arrival (TOAs) of pulses from pulsars. Such measurements can be used for a variety of scientific purposes, including characterizing energy loss from gravitational-wave emission in binary systems \citep[e.g.,][]{wnt10}, testing other predictions of the general theory of relativity \citep{stairs03}, detecting small bodies orbiting pulsars \citep[e.g.,][]{wolszczan94}, and studying the structure of the ionized interstellar medium \citep[e.g.,][]{ne2001}. Additionally, current efforts by pulsar timing arrays (PTAs; \citealt{dcl+16,hobbs13}), including the North American Nanohertz Observatory for Gravitational Waves (NANOGrav; \citealt{11yr-timing}), seek to use precision pulsar timing to directly detect nanohertz gravitational radiation.

All applications of pulsar timing rely on the  link between arrival times and the spin phase of a neutron star (NS) and the spin stability of the NS itself. While past successes have shown that pulsars make excellent clocks, they are not perfect in this respect, and understanding their limitations is crucial to all pulsar timing efforts. Such limitations arise from a number of sources, including rapid spin-up events called glitches \citep{elsk11}, which are thought to be a result of differential rotation between a neutron star's crust and its superfluid interior \citep{ga09}; and torque fluctuations in the magnetosphere, which produce spin noise with power concentrated at low frequencies \citep{klo+06,lhk+10}. Additionally, 
the radio emission from a pulsar is intrinsically variable, causing the shapes of individual pulses to vary stochastically. This variability is likely the result of changes in the relativistic flows in the magnetosphere. Fortunately, averages of many pulses converge to shapes that have proven to be stable on timescales of decades, albeit with remaining low-level stochastic shape variations.  

Pulse shape stability is the basis for the template matching algorithms used to convert pulse shapes into TOA measurements \citep{taylor92,cordes13a}. For the most part, these algorithms rely on a simple model in which an observed pulse profile, constructed as the average of a large number of pulses, is treated as having a known, fixed shape, modified only by the addition of white noise. When the profile deviates from this template shape, TOA estimation errors result. Such deviations occur in practice for several reasons, many of which our outlined in Section~\ref{sec:causes} below. In particular, since a profile is the sum of only finitely many pulses, and the shapes of individual pulses vary stochastically, the shape of the profile also varies, albeit at a lower level. This single-pulse stochasticity limits the precision of TOA estimates for pulsars observed at very high signal-to-noise ratio (S/N).

In this paper we describe the TOA estimation errors that arise from various kinds of pulse shape changes, and describe methods for identifying the presence of such shape changes and ascertaining the extent to which they affect TOA measurements. In Section~\ref{sec:amsn}, we describe the amplitude-modulated shot noise model underlying our current understanding of pulsar emission. In Section~\ref{sec:toa-estimation}, we introduce the conventional algorithm used for TOA generation, and describe how the corresponding uncertainties may be estimated. Formulas for predicting the TOA estimation errors caused by particular shape changes are introduced and applied to simulated data in Section~\ref{sec:jitter-noise}. A series of techniques for characterizing pulse shape changes, particularly those created by single-pulse stochasticity, and measuring the TOA estimation errors they create, is outlined in Section~\ref{sec:sp-stochastic}. Sections~\ref{sec:state-changes} and~\ref{sec:rfi} describe other phenomena which may cause shape changes, and outline how the previously-described methods may be applied to them. A subsequent paper will discuss methods for mitigating the effects of TOA estimation errors caused by pulse shape changes in certain cases.

\section{Causes of Pulse Shape Changes}\label{sec:causes}

Observed pulse shapes deviate from the template pulse shape for several reasons. Here, we will primarily be concerned with pulse jitter; i.e., shape changes caused by the intrinsic stochastic behavior of single pulses \citep{cd85,cordes90,lkl+12}. This should be understood to include variability in the amplitude and shape of pulse components, as well as their phases.  Jitter is generally broadband, i.e., strongly correlated between adjacent frequencies, and originates in the emission region, possibly as the result of randomness in the excitation of plasma modes at the polar cap pair formation front~(e.g., \citealt{pts20}). We describe several ways to characterize timing errors caused by jitter, with reference to simulations and data from observations of the Vela pulsar, PSR B0833$-$45, in Section~\ref{sec:sp-stochastic}.

Some pulsars also show other kinds of intrinsic pulse shape changes. Nulling pulsars \citep{backer70a,gjk12,sm21}  occasionally cease to emit detectable pulses for short periods of time, ranging from a few pulse periods to several days. Mode-changing pulsars, including PSR B0329+54 and PSR B1237+25 \citep{backer70b,lyne71,bmsh82} alternate between two or more modes of emission on similar time scales. In at least one millisecond pulsar, PSR J1643$-$1224, evidence has been seen for a shape change that appears abruptly, decays over a period of months, and results in a permanently modified pulse shape~\citep{slk+16}. A similarly abrupt shape change, which may have a similar origin, was recently seen in the millisecond pulsar J1713+0747~\citep{xhb+21,lam21}.
Approaches to characterizing these phenomena and their effects on timing will be described in Section~\ref{sec:state-changes}.

Radio emission from pulsars is dispersed and scattered as it travels through the interstellar medium (ISM). This, too, causes pulse shape changes~\citep[see, e.g.,][]{rickett90}. The most significant ISM propagation effect is dispersion, which creates a delay proportional to the inverse square of the observation frequency, $\nu$, and to the column density of electrons along the line of sight, known in this context as the dispersion measure (DM). Another significant effect is interstellar scattering, a result of the propagation of radio waves along multiple paths through the nonuniform distribution of electrons in the ISM. This leads to a frequency-dependent broadening of pulse profiles, as well as to delays proportional to $\nu^{-4.4}$ (assuming density fluctuations in the ISM can be described by a Kolmogorov spectrum). The detailed analysis of these phenomena and their effects on pulsar timing is closely tied to the physics and astrophysics of the interstellar medium, and is beyond the scope of this paper.

Instrumental effects and radio-frequency interference (RFI) can also produce effective pulse shape changes. Because pulsar emission is often strongly polarized, imperfections in the polarization response and calibration of a receiver can change the apparent pulse shape unless they are carefully accounted for \citep{hpn+01,12yr-polarimetry}. Additionally, channel gain and timing mismatches in time-interleaved analog-to-digital converter (ADC) systems can lead to image rejection artefacts that present themselves as frequency-reversed ``ghost images'' of the dispersed pulse~(\citealt{12yr-nb-timing}, \S2.3.1; see also \citealt{kkm+01}). The most significant effect, however, is generally RFI. This is often manifest as short impulses, which can alter the shape of an observed pulse if they coincide with it in time, or as long-term, narrow-band signals, which, if not removed from the data, can create a sinusoidal ripple effect most noticeable in the off-pulse region of a profile. Some of the techniques described here lend themselves well to identifying data which has been corrupted by RFI and understanding the extent to which it affects pulsar timing. This will be explored further in Section~\ref{sec:rfi}.

\section{The Amplitude-Modulated Shot Noise Model}\label{sec:amsn}


\smallskip

At very high time resolution, radio emission from pulsars consists of polarized shot noise~\citep{cordes76,jap01,cbh+04}. That is, the electric field, $\vec{E}(t)$, at the antenna is given by a superposition of coherent shot pulses:
$\vec{E}(t)=\sum_j\vec{e}_j(t-t_j)$,
where each $\vec{e}_j(t-t_j)$ is a shot arriving at time $t_j$. Shots can be assumed to occur as a Poisson process in time, with rate $\eta(t)$. The Fourier transform of the shot shape determines the spectrum of the pulsar, so the fact that pulsars are detected at radio frequencies above \SI{1}{GHz} means that the intrinsic width of a typical shot is less than \SI{1}{ns}.

When observed with a quadrature heterodyned receiver system, the pulsar signal is shifted to baseband by mixing it with a local oscillator signal, and multiplicaton by a lowpass 
filter with impulse response $h(t)$. The result is then sampled to produce a baseband voltage time series, $\vec{\mc{E}}(t)$, with two complex-valued components, one for each polarization:
\begin{equation}
\vec{\mc{E}}(t) = \int_{-\infty}^{\infty} \vec{E}(t')\mkern2mu\ee^{-2\pi\ii\nu_0 t'} h(t-t')\dd{t'}.
\end{equation}
The observed baseband time series, $\vec{\mc{E}}(t)$, still consists of shots, but the shape of each shot is modified by convolution with the bandpass response, $h(t)$, i.e., we have
$\vec{\mc{E}}(t) = \sum_j \vec{s}_j(t-t_j)$,
where
\begin{equation}
\vec{s}_j(t) = \int_{-\infty}^{\infty}\vec{e}_j(t')\mkern2mu \ee^{-2\pi\ii\nu_0 t'} h(t-t')\dd{t'}.
\end{equation}
The width of the bandpass response, $h(t)$, is approximately equal to the reciprocal of the bandwidth, which is usually significantly larger than the intrinsic width of the shots. (For modern wideband receiver systems with bandwidths on the order of hundreds of MHz to a few GHz, this is less true, but data from such systems are often split into channels with widths of order MHz for analysis.)

From Campbell's theorem~\citep[cf.][]{rice44}, it follows that the ensemble average signal amplitude, or envelope function, $A(t)$, is given by
\begin{equation}\label{eqn:envelope-fn}
A(t) = \bracket{\abs{\vec{\mc{E}}(t)}^2} = \int_{-\infty}^{\infty} \bracket[big]{\abs{\vec{s}(t-t')}^2}\mkern2mu \eta(t')\dd{t'},
\end{equation}
where $\eta(t)$ is the shot rate. We can therefore write
\begin{equation}
\vec{\mc{E}}(t) = a(t) \mkern1mu \vec{m}(t),
\end{equation}
where $a(t)^2 = A(t)$, and $\vec{m}(t)$ is a noise process with unit variance, i.e., $\bracket*{\abs{\vec{m}(t)}^2} = 1$. The total intensity, $I(t) = \abs{\vec{\mc{E}}(t)}^2$, is then given by
\begin{equation}
I(t) = A(t) M(t),
\end{equation}
where $M(t)=\abs{\vec{m}(t)}^2$. When $\eta(t)$ becomes sufficiently large, the distribution of $\vec{m}(t)$ becomes approximately Gaussian, and that of $M(t)$ approaches a chi-squared distribution, as in the simple amplitude-modulated noise model of \citet{rickett75}.

The stochastic behavior of $M(t)$ leads to so-called ``self noise'', a form of pulse shape variability that is present even when $A(t)$ is constant in time. However, for the remainder of this paper, we will be concerned primarily with stochastic variations in the envelope function $A(t)$. When such variations are present, the expectation in equation~(\ref{eqn:envelope-fn}) should be understood as averaging over variations in the arrival time, amplitude, and shape of individual shots, but not over variations in the envelope function itself.

\section{TOA Estimation and Errors from Additive Noise}\label{sec:toa-estimation}

\noindent The conventional matched-filtering algorithm used for TOA estimation is laid out in detail in \citet[Appendix A]{taylor92}. It can be characterized as the maximum likelihood estimator for the model in which the profile, $p(\phi)$, is described as
\begin{equation}\label{eqn:profile-model-cp}
    p(\phi)=a u(\phi-\tau) + n(\phi),
\end{equation}
where $u(\phi)$ is the template (normalized to unit maximum), $n(\phi)$ is white noise with variance $\sigma_n^2$, and the amplitude $a$ and phase offset $\tau$ are the parameters. Both $p(\phi)$ and $u(\phi)$ should be thought of as periodic in $\phi$ with period $1$. We are using a continuous notation for clarity here -- in reality, values of $p(\phi)$ and $u(\phi)$ are known only at a finite number, $N_\phi$, of phase bins. Shifting $u(\phi)$ by a fractional number of phase bins can be performed by operating in the frequency domain and making use of the Fourier shift theorem.

The likelihood for this model has the form $\mc{L}\of{a,\tau}\propto\ee^{-\frac12\chi^2\of{a,\tau}}$, where
\begin{equation}\begin{split}\label{eqn:chisquared-cp}
    \chi^2\of{a,\tau} &= \frac{N_\phi}{\sigma_n^2}\paren{\int_0^1 p(\phi)^2 \dd\phi - 2a \int_0^1 p(\phi) u(\phi-\tau) \dd\phi + a^2\int_0^1 u(\phi)^2 \dd\phi}.
\end{split}\end{equation}
Completing the square in $a$ gives
\begin{equation}
    \chi^2\of{a,\tau}=\frac{N_\phi \bracket{u^2}}{\sigma_n^2}\sbrack{\paren{a-\hat{a}(\tau)}^2-\hat{a}(\tau)^2}+\frac{N_\phi\bracket{p^2}}{\sigma_n^2},
\end{equation}
where $\bracket{u^2}=\int_0^1 u(\phi)^2 \dd\phi$, $\bracket{p^2}=\int_0^1 p(\phi)^2 \dd\phi$ and
\begin{equation}\label{eqn:ahat}
    \hat{a}(\tau)=\frac1{\bracket{u^2}}\int_0^1 p(\phi) u(\phi-\tau) \dd\phi.
\end{equation}
It follows that the maximum likelihood (minimum $\chi^2$) occurs at $\tau=\hat\tau$ and $a=\hat{a}(\hat\tau)$, where
\begin{equation}\label{eqn:max-problem}
    \hat{\tau}=\operatorname*{argmax}_\tau\hat{a}(\tau)^2 = \operatorname*{argmax}_\tau\hat{a}(\tau).
\end{equation}
The quantity $\hat{a}(\tau)$ is, up to a constant factor, the cross-correlation between the profile and the template. The time-of-arrival estimate, $\hat\tau_{\mr{MF}}$, that maximizes the likelihood of model~(\ref{eqn:profile-model-cp}) can therefore be computed by maximizing $\hat{a}(\tau)$ numerically.
This procedure is identical to that described by \citet{taylor92}.

The uncertainties in $a$ and $\tau$ may be determined by expanding $\chi^2$ in a power series around its minimum. By calculating the appropriate derivatives, one can determine that
\begin{equation}
    \chi^2(a,\tau) \approx \chi^2(\hat{a},\hat\tau) + \frac{(a-\hat{a})^2}{\sigma_a^2} +  \frac{(\tau-\hat\tau)^2}{\sigma_\tau^2},
\end{equation}
where the variances, $\sigma_a^2$ and $\sigma_\tau^2$, are given by
\begin{align}
    \label{eqn:sigma-a}
    \sigma^2_{a}&=\frac{\sigma_n^2}{N_\phi\bracket{u^2}},\\
    \label{eqn:sigma-tau}
    \sigma^2_{\tau}&=\frac{\sigma_n^2}{\hat{a}N_\phi}\paren{-\int_0^1 p(\phi) u''(\phi-\hat\tau) \dd\phi}^{-1}.
\end{align}
If the profile is well described by the model (equation~\ref{eqn:profile-model-cp}), one can substitute $p(\phi)=\hat{a}u(\phi-\hat\tau)$ in equation (\ref{eqn:sigma-tau}) and integrate by parts to obtain
\begin{equation}\label{eqn:err-from-weff}
    \sigma_\tau^2=\frac{W_{\mr{eff}}^2}{S^2N_\phi},
\end{equation}
where
\begin{equation}\label{eqn:weff}
    W_{\mr{eff}}=\paren{\int_0^1 u'(\phi)^2\dd\phi}^{-1/2}
\end{equation}
is a quantity with units of phase which can be interpreted as an effective pulse width, and $S=\hat{a}/\sigma_n$ is the signal-to-noise ratio of the profile.

\section{Arrival time errors from single-pulse stochasticity}
\label{sec:jitter-noise}


All radio pulsars produce single pulses with stochastic variations in their amplitude, shape, and pulse phase. An average profile comprising a large but finite number of pulses ($N$) will therefore deviate from the assumed template shape by an amount proportional to $N^{-1/2}$. This deviation in shape produces an error in the corresponding arrival time estimate. Below, we use the term ``jitter'' to refer to these single pulse variations, and ``jitter noise'' to refer to their effect on TOA estimates.

In this section we develop a pulse and timing model that captures the salient features of jitter.    

\noindent Suppose that the profile, $p(\phi)$, differs from the shifted, scaled template by a small amount. That is,
\begin{equation}\label{eqn:residual}
  p(\phi) = a u(\phi-\tau) + r(\phi),
\end{equation}
where the profile residual, $r(\phi)$, is much smaller in magnitude than $a u(\phi-\tau)$. The presence of $r(\phi)$ means that the matched-filtering estimate of the phase shift, $\hat\tau$, will differ from the true phase shift by a small amount, $\delta\tau$. As long as $\delta\tau$ is small compared to the width of the template, the maximization in equation~(\ref{eqn:max-problem}) can be carried out analytically, by expanding the condition $\hat{a}'\of{\tau+\delta\tau} = 0$ to first order in $\delta\tau$ and solving for $\delta\tau$. This gives
\begin{equation}\label{eqn:delta-tau-basic}
    \delta\tau\approx\frac{\int_0^1 p(\phi)u'(\phi-\tau)\dd\phi}{\int_0^1 p(\phi) u''(\phi-\tau)\dd\phi}.
\end{equation}
Making use of equation~(\ref{eqn:residual}), we can rewrite this as
\begin{equation}
    \delta\tau\approx\frac{a\int_0^1 u(\phi-\tau)u'(\phi-\tau)\dd\phi + \int_0^1 r(\phi)u'(\phi-\tau)\dd\phi}{a\int_0^1 u\of{\phi-\tau} u''\of{\phi-\tau}\dd\phi + \int_0^1 r(\phi)u''(\phi-\tau)\dd\phi}.
\end{equation}
The first term in the numerator is the integral of a total derivative and so vanishes. Furthermore, since $r\of\phi\ll a u\of{\phi-\tau}$, the second term in the denominator can be neglected compared to the first. The above expression for $\delta\tau$ can therefore be simplified to
\begin{equation}\label{eqn:projection}
    \delta\tau\approx-\frac{\int_0^1 r(\phi) \mkern1mu u'(\phi-\tau)\dd\phi}{a\int_0^1 u'(\phi-\tau)^2\dd\phi}.
\end{equation}
In other words, the error in the matched-filtering estimate of $\tau$ is proportional to the projection of $r(\phi)$ onto $u'(\phi-\tau)$.

Importantly, while equation~(\ref{eqn:projection}) is useful for estimating $\delta\tau$ when $\tau$ is known, it cannot be used directly to make an improved estimate of $\tau$ when only $\hat\tau$ is known. In such a case, $r(\tau)$ may be estimated as
\begin{equation}\label{eqn:estimated-residual}
\hat{r}(\phi) = p(\phi) - \hat{a}(\hat\tau) u(\phi-\hat\tau).
\end{equation}
Solving for $p(\phi)$ and substituting the result into equation~(\ref{eqn:delta-tau-basic}), replacing $\tau$ with $\hat\tau$ where appropriate, would appear to give the following estimate of $\delta\tau$:
\begin{equation}\label{eqn:delta-tauhat}
\delta\hat\tau = -\frac{\int_0^1 \hat{r}(\phi) \mkern1mu u'(\phi-\hat\tau)\dd\phi}{\hat{a}(\hat\tau)\int_0^1 u'(\phi-\hat\tau)^2\dd\phi}.
\end{equation}
However, as a result of the way it is constructed, $\hat{r}(\phi)$ is approximately orthogonal to $u'(\phi-\hat\tau)$, as may be determined by expanding the numerator of equation~(\ref{eqn:delta-tauhat}) to first order in $\delta\tau$. In particular, we have
\begin{equation}
\int_0^1 \hat{r}(\phi) \mkern1mu u'(\phi-\hat\tau)\dd\phi = 0 + \mc{O}(\delta\tau^2),
\end{equation}
so equation~(\ref{eqn:delta-tauhat}) becomes $\delta\hat\tau = 0 + \mc{O}(\delta\tau^2)$. This means that $\delta\hat\tau$ is not a useful estimator of $\delta\tau$.

\subsection{Component Amplitude and Phase Variations}

\begin{figure}
\centering
\includegraphics[width=0.33\textwidth]{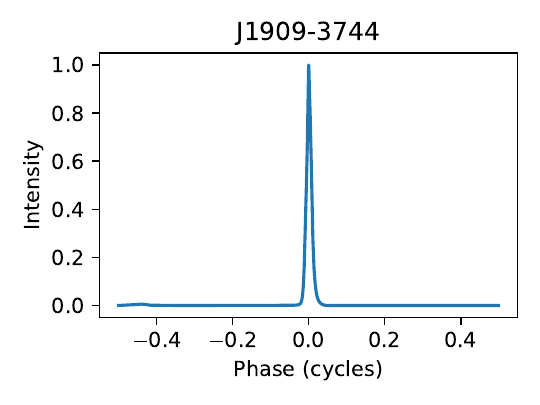}\hspace{-1em}
\includegraphics[width=0.33\textwidth]{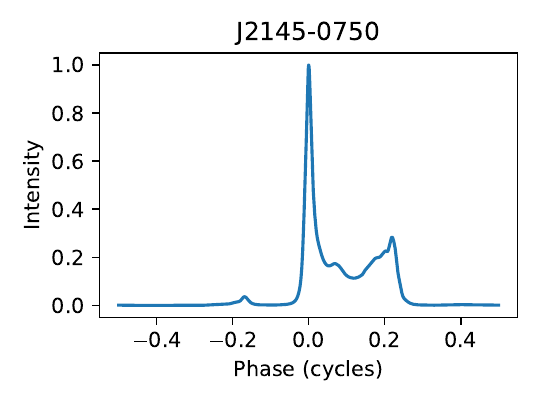}
\includegraphics[width=0.33\textwidth]{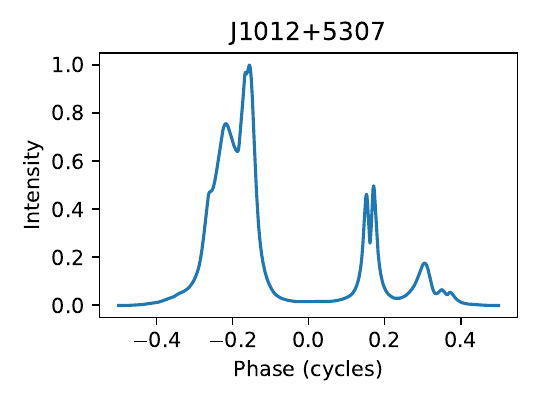}
\caption{
Average profiles for three millisecond pulsars (PSRs J1909$-$3744, J2145$-$0750, and J1012$+$5307), demonstrating the variety of pulse shapes exhibited by MSPs. All three profiles were taken from the NANOGrav 12.5-year data release \citep{12yr-nb-timing}, and are derived from observations made using the 1--2 GHz receiver and GUPPI backend on the Green Bank Telescope between 2010 and 2017.}
\label{fig:msp-profiles}
\end{figure}

The average profiles of pulses from pulsars, especially millisecond pulsars (MSPs), are often complex and multi-peaked, as seen in Figure~\ref{fig:msp-profiles}, with components of single pulses often differing in amplitude and phase from those seen in the average profile. To model this, we can assume that the pulse is made up of several components of fixed shape, each of which varies in amplitude and phase. The profile is then given by
\begin{equation}\label{eqn:profile-components}
    p(\phi) = \frac{a}{N}\sum_{i=1}^N\sum_{j=1}^{N_c} \paren{1+b_{ij}}\mkern1mu c_{j}(\phi-\psi_{ij}),
\end{equation}
where $N$ is the number of pulses being averaged, $N_c$ is the number of components, $c_{j}(\phi)$ is the shape of component $j$, and $b_{ij}$ and $\psi_{ij}$ are the amplitude and phase offsets, respectively, of component $j$ in pulse $i$. Without loss of generality, we can take the ensemble average amplitude and phase offsets for each component to be zero. Equation~(\ref{eqn:profile-components}) can then be expanded to first order in $\psi_j$, giving
\begin{equation}\label{eqn:profile-expansion}
    p(\phi) \approx \frac{a}{N}\sum_{i=1}^N\sum_{j=1}^{N_c}\sbrack{\paren{1 + b_{ij}} c_j(\phi) - \paren{\psi_{ij} + b_{ij}\psi_{ij}} c_j'(\phi)}.
\end{equation}
The template is the normalized ensemble average profile:
\begin{equation}\label{eqn:template}
    u(\phi)=\frac{\bracket{p(\phi)}}{a}\approx\sum_{j=1}^{N_c}c_j(\phi).
\end{equation}

The profile residual, $r(\phi)=p(\phi)-a u(\phi)$, is therefore
\begin{equation}\label{eqn:model-residual}
    r(\phi)\approx a\sum_{j=1}^{N_c}\sbrack{\overline{b_j} c_j(\phi) + \paren[big]{\overline{\psi_j} + \overline{b_j\psi_j}} c_j'(\phi)},
\end{equation}
where barred quantities represent sample averages. The first term in equation~(\ref{eqn:model-residual}) arises from the difference in amplitude between the profile and the template, and the second from the difference in phase. Substituting this into equation~(\ref{eqn:projection}) gives
\begin{equation}\label{eqn:tauhat-full}
    \delta\tau\approx\frac{\sum_{j=1}^{N_c} \sbrack{\paren[big]{\overline{\psi_j} + \overline{b_j\psi_j}} \Gamma_j + \overline{b_j}\Delta_j}} {\sum_{j=1}^{N_c} \Gamma_j},
\end{equation}
where we have made use of the symbols
\begin{align}
    \Gamma_{j} &= \sum_{k=1}^{N_c}\int_0^1  c_j'(\phi)\mkern2mu c_k'(\phi)\dd\phi\quad\text{and}\label{eqn:Gamma}\\
    \Delta_{j} &= \sum_{k=1}^{N_c}\int_0^1  c_j(\phi)\mkern2mu c_k'(\phi)\dd\phi\label{eqn:Delta}
\end{align}
for cross-correlations between the component shapes, and ignored all but the leading term in the denominator.
When there is only one component, equation~(\ref{eqn:tauhat-full}) reduces to
\begin{equation}
    \hat\tau\approx\overline{\psi}+\overline{b\psi}.
\end{equation}
The variance of the TOA estimate is therefore
\begin{equation}\label{eqn:toa-variance-1comp}
    \sigma_\tau^2=\bracket[big]{\hat\tau^2}=\frac{1+m^2}{N}\bracket[big]{\psi^2},
\end{equation}
where $m^2=\bracket{b^2}$ is the square of the single-pulse modulation index. This shows that, for a single-component pulse, phase variation alone can produce TOA errors. Amplitude variation may enhance TOA errors created by phase variation, but cannot produce them on its own.

For an arbitrary number of components,  the variance of the TOA estimate from equation~(\ref{eqn:tauhat-full}) is
\begin{equation}\label{eqn:tauhat-variance}
    \sigma_\tau^2=\frac{\sum_{j=1}^{N_c} \sbrack{\paren{1+m_j^2}\bracket{\psi_j^2} \Gamma_j^2 + m_j^2 \Delta_j^2}}{N\paren[big]{\sum_{j=1}^{N_c} \Gamma_j}^2},
\end{equation}
where $m_j^2=\bracket{b_j^2}$ is the square of the single-pulse modulation index for component $j$. Unlike in the single component case, here amplitude variation can create TOA errors on its own, even in the absence of phase variation: setting $\bracket{\psi_j}^2=0$ in equation~(\ref{eqn:tauhat-variance}) leaves the term involving $\Delta_j^2$. Conceptually, this is because amplitude variations represent true shape changes only when there are multiple components with separately varying amplitudes. For single component pulses, amplitude variations simply scale the pulse, and can be absorbed into the phase factor, $a$, but for pulses with multiple components that vary separately, this is no longer the case.

\subsection{Simulations with Gaussian Components}

\begin{figure}
\centering
\includegraphics[width=0.7\textwidth]{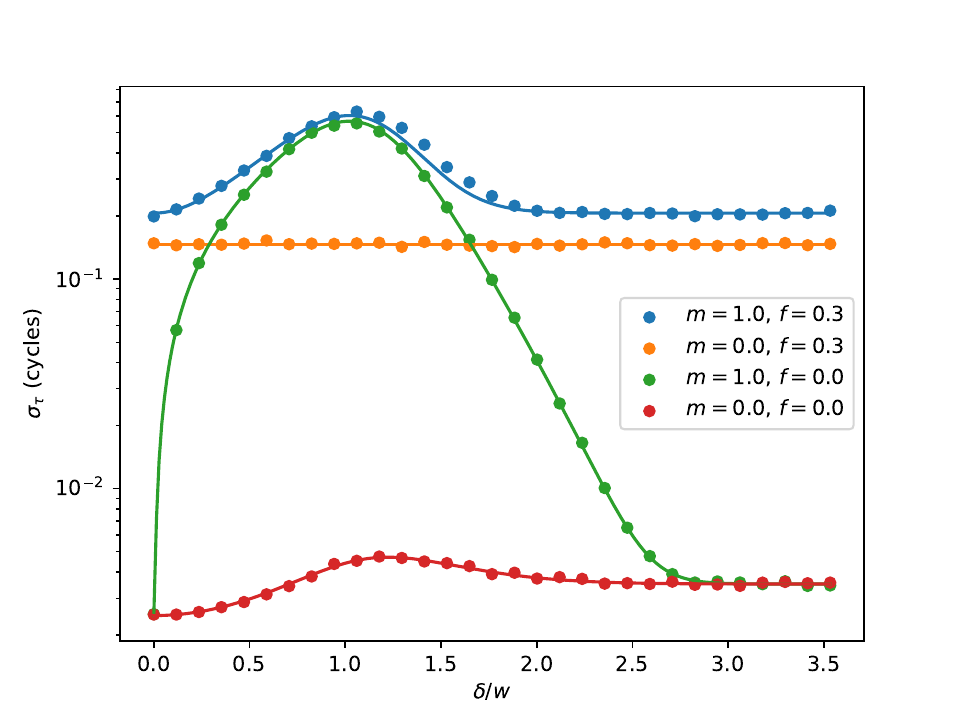}
\caption{
The standard deviation of TOA estimates in simulated profiles with two identical components of width $w=0.0425$ and varying spacing. Crosses indicate results of simulated cases, and solid lines indicate predictions based on equations~(\ref{eqn:err-from-weff}) and~(\ref{eqn:twocomp-prediction}). Four cases are shown: one in which the pulses have amplitude and phase variations ($m=1.0$, $f=0.3$ for each component), one in which the pulses have only phase variations ($m=0.0$, $f=0.3$), one in which they have only amplitude variations ($m=1.0$, $f=0.3$), and a reference case where the pulses are copies of the template with additive white noise ($m=0.0$, $f=0.0$). Each point was calculated based on $N_p=2048$ profiles, each the average of $N=1000$ pulses. The average profiles had a signal-to-noise ratio $S=1000$.}
\label{fig:2comps-sep}
\end{figure}

\begin{figure}
\centering
\includegraphics[width=\textwidth]{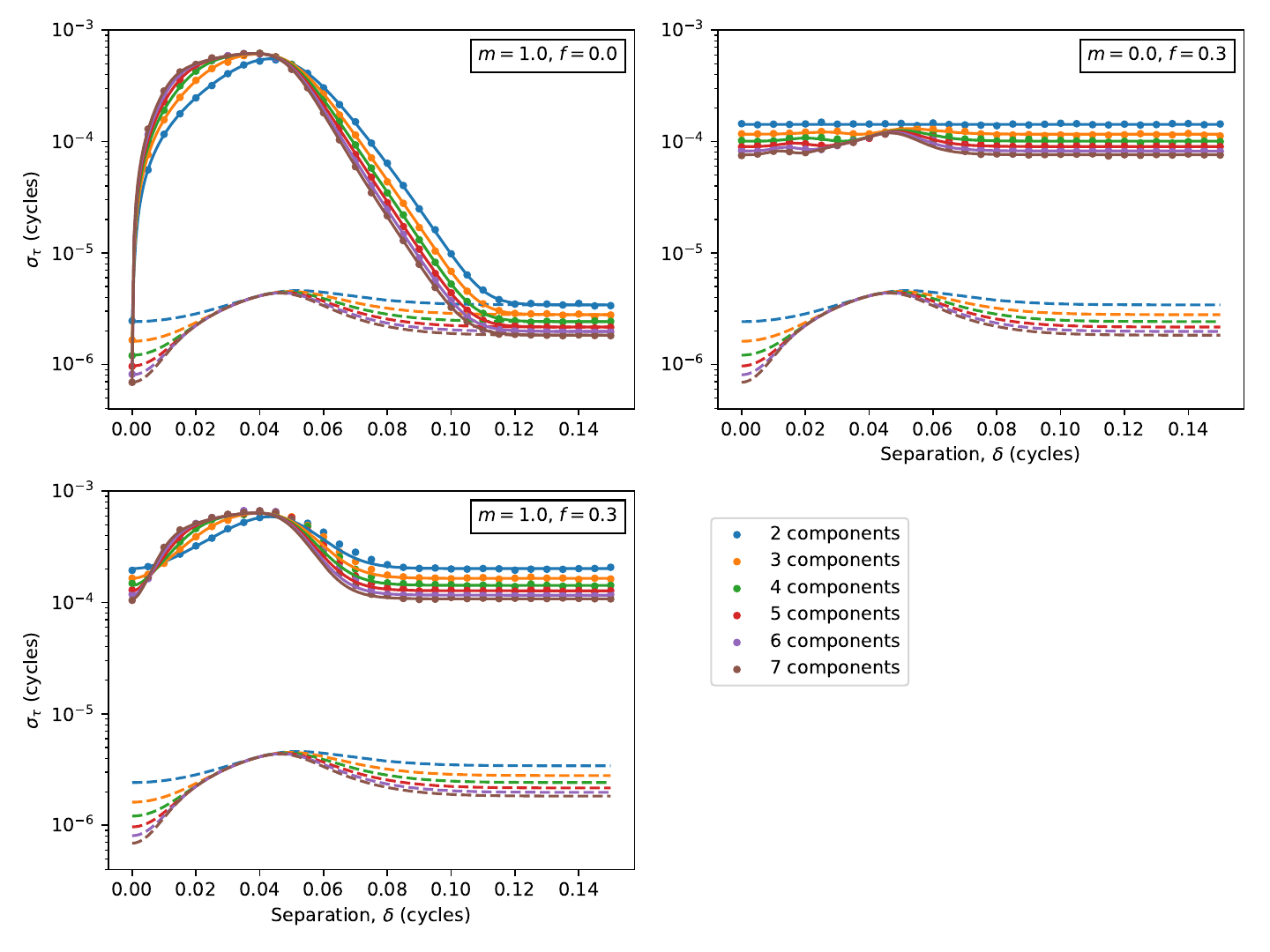}
\caption{The standard deviation of TOA estimates in simulated profiles with several identical components of width $w=0.0425$ and varying spacing. Results with amplitude variations only ($m=1.0$) is shown in the upper left, results with phase variations only ($f_j=0.3$) in the upper right, and results with both amplitude and phase variations ($m=1.0$, $f_j=0.3$) in the lower left. As in Figure~\ref{fig:2comps-sep}, filled circles indicate simulation results, and solid lines indicate predictions based on equation~(\ref{eqn:tauhat-variance}). For comparison, the predictions for the reference case with no amplitude or phase variations (based on equation~\ref{eqn:err-from-weff}) are shown as dashed lines.}
\label{fig:ncomps-sep}
\end{figure}

We performed a number of simulations in which shape variations were introduced into a simple pulse shape model consisting of Gaussian components. In each simulation, $N_p$ profiles were generated by averaging $N$ simulated pulses each, in the manner of equation~(\ref{eqn:profile-components}). The shape of each component was taken to be Gaussian, with amplitude $a_j$ and width $w_j$:
\begin{equation}
c_j(\phi) = a_j\exp\sbrack{-\frac{(\phi-\mu_j)^2}{2w_j^2}}.
\end{equation}
With this choice of component shape, we have
\begin{align}
	\Gamma_j &= \sqrt{2\pi}a_jw_j\sum_{k=1}^{N_c}\frac{a_k w_k}{\paren{w_j^2+w_k^2}^{\frac32}}\sbrack{1-\frac{(\mu_j-\mu_k)^2}{w_j^2+w_k^2}}\exp\sbrack{-\frac{\paren{\mu_j-\mu_k}^2}{2\paren{w_j^2+w_k^2}}}\label{eqn:Gamma-gaussian}\\
	\Delta_j &= \sqrt{2\pi}a_jw_j\sum_{k=1}^{N_c}\frac{a_k w_k\paren{\mu_j-\mu_k}}{\paren{w_j^2+w_k^2}^{\frac32}}\exp\sbrack{-\frac{\paren{\mu_j-\mu_k}^2}{2\paren{w_j^2+w_k^2}}}.\label{eqn:Delta-gaussian}
\end{align}
Amplitudes ($1+b_{ij}$) were drawn from a log-normal distribution with modulation index $m_j$, and phase offsets ($\psi_{ij}$) were drawn from a normal distribution with standard deviation $\bracket[big]{\psi_j^2} = f_jw_j$. The ratio $f_j=\bracket[big]{\psi_j^2}/w_j$, where $w_j$, as here, is the single-pulse width of a particular component, is called the jitter parameter. Below, we will distinguish between different simulated cases by specifying the amplitudes and widths of each component, the separation between components, and the modulation index, $m_j$ and jitter parameter, $f_j$, of each component.

As an initial comparision of simulations with theoretical results, consider the case where pulses are made up of two identical components. 
In this case, we have $\Gamma_1 = \Gamma_2 = \Gamma$ and $\Delta_1 = -\Delta_2 = \Delta$, where
\begin{align}
	\Gamma &= \frac{\sqrt{\pi}a^2}{2w}\sbrack{1+\paren{1-\frac{\delta^2}{2w^2}}\exp\of{-\frac{\delta^2}{4w^2}}},\\
	\Delta &= \frac{\sqrt{\pi}a^2\delta}{2w}\exp\of{-\frac{\delta^2}{4w^2}},
\end{align}
and $\delta=\mu_1-\mu_2$ is the separation between the components. Equation~(\ref{eqn:tauhat-variance}) then predicts that the variance of the TOA estimates will be given by
\begin{equation}\begin{split}\label{eqn:twocomp-prediction}
	\sigma_\tau^2 &= \frac{(1+m^2)\bracket{\psi^2}}{2N} + \frac{m^2\Delta^2}{2N\Gamma^2}\\
	 &= \frac{(1+m^2)f^2w^2}{2N} + \frac{m^2\delta^2}{2N}\sbrack{\exp\of{\frac{\delta^2}{4w^2}}+\paren{1-\frac{\delta^2}{2w^2}}}^{-2}.
\end{split}\end{equation}
Figure~\ref{fig:2comps-sep} compares the predictions of equation~(\ref{eqn:twocomp-prediction}) with simulation results for various values of $\delta$ in three different cases: one with only amplitude variations, another with only phase variations, and a third with both types of variation. With only phase variations, $\sigma_\tau$ is independent of $\delta$, while in cases where phase variations are included, it reaches a maximum when $\delta\approx w$. For reference, a fourth case with neither amplitude nor phase variations, but only a small amount of additive white noise, is shown. In this case, $\sigma_\tau$ is predicted by equation~(\ref{eqn:err-from-weff}). It has some dependence on $\delta$ because $W_{\mr{eff}}$ is a function of the template shape (equation~\ref{eqn:weff}).

A more extensive set of simulation results is shown in Figure~\ref{fig:ncomps-sep}. Shown there are the results of simulations with various numbers of identical components (between 2 and 7), in each of the three primary cases from Figure~\ref{fig:2comps-sep}, along with the predictions of equation~(\ref{eqn:tauhat-variance}). One notable trend visible in Figure~\ref{fig:ncomps-sep} is that, for large separations ($\delta\gg w$), $\sigma_\tau$ decreases as the number of components, $N_c$, is increased. Indeed, it is approximately proportional to $N_c^{-1/2}$. This happens because, for large values of $\delta$, the cross-correlation $\Delta_j$ rapidly approaches $0$, as do all terms in equation~(\ref{eqn:Gamma}) for $\Gamma_j$ except the $j=k$ term, which is independent of $\delta$. Using equation~(\ref{eqn:Gamma-gaussian}), we can see that, for the identical Gaussian components used in the simulations, $\Gamma_j$ approaches $\sqrt{\pi}a^2/(2w)$ for each component. It follows that, in the limit of large $\delta$, we have
\begin{equation}
\sigma_\tau^2 = \frac{(1+m^2)f^2w^2}{NN_c}.
\end{equation}
A similar phenomenon occurs when the components are not identical, complicated only by the fact that the components do not contribute equally to the variance. In general, for sufficiently well-separated components, we have
\begin{equation}\label{eqn:independent-components}
\sigma_\tau^2 = \frac{\sum_{j=1}^{N_c}a_j^4(1+m_j^2)f_j^2}{N\paren[big]{\sum_{j=1}^{N_c}a_j^2w_j^{-1}}^2}.
\end{equation}
In other words, the TOA estimation errors associated with each component combine in a manner weighted by the combination $a_j^2 w_j^{-1}$, and tending to decrease as the number of components increases. However, the trend of decreasing $\sigma_\tau$ with increasing $N_c$ is not universal. As seen in Figure~\ref{fig:ncomps-sep}, in some cases with amplitude variations included, when $\delta\lesssim w$, $\sigma_\tau$ actually increases with increasing $N_c$ at a similar rate. In combination with the fact that $\sigma_\tau$ also depends significantly on the modulation index, $m$, and jitter parameter, $f$, this means that it is generally not possible to predict $\sigma_\tau$ from the number of pulse components without further information.

\section{Tools for Assessing Single-Pulse Stochasticity}\label{sec:sp-stochastic}
Several statistical tools can be used to identify the effects of single-pulse stochasticity in pulsar observations, and assess the degree to which these effects influence TOA estimates based on the observations. One can estimate the effect of single-pulse stochasticity on timing by comparing the standard deviation of TOA residuals to the signal-to-noise ratio and number of pulses averaged in the profiles used to generate them. The phase autocorrelation functions of single pulses and average profiles can be used to establish that their widths differ. Correlations between TOAs in adjacent frequency bands can be used to separate timing noise caused by single-pulse stochasticity from other forms of noise. And principal component analysis (PCA) allows the dominant modes of shape variation to be identified. Below, we use each of these tools to compare the results of simulations to observations of the Vela pulsar, PSR B0833$-$45. First, however, we will describe the particular simulations and observations under consideration.

\subsection{TOA and Mismatch Statistics}

\begin{figure}
\centering
\includegraphics[width=0.7\textwidth]{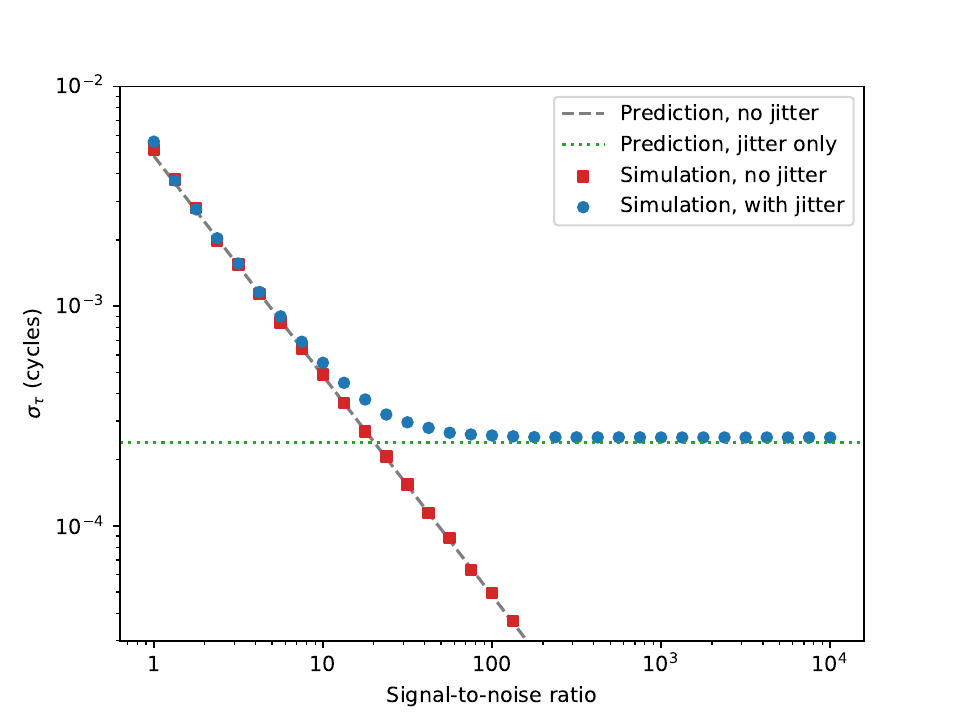}
\caption{The standard deviation of TOA estimates for simulated pulses with amplitude and phase jitter and varying signal-to-noise ratio. Each point was calculated based on $N_p=2048$ simulated profiles, each the average of $N=1000$ simulated pulses. The dashed gray line shows the predicted TOA error based on equation~(\ref{eqn:err-from-weff}) alone, ignoring jitter, and the dotted green line shows the predicted TOA error from jitter alone, based on equation~(\ref{eqn:tauhat-variance}).}
\label{fig:jitter-dominated-regime}
\end{figure}

The TOA error caused by jitter becomes larger than the error from radiometer noise when the signal-to-noise ratio $S$, measured as the ratio of peak flux to off-pulse RMS, exceeds a threshold value determined by the pulse duty cycle. One can therefore measure its magnitude by comparing the standard deviation, $\sigma_\tau$, of a set of TOA residuals to the corresponding signal-to-noise ratio, as shown in Figure~\ref{fig:jitter-dominated-regime}. At lower signal-to-noise ratios, $\sigma_\tau$ is proportional to $S^{-1/2}$, a consequence of equation~(\ref{eqn:err-from-weff}), whereas at higher signal-to-noise ratios, $\sigma_\tau$ becomes independent of $S$, approaching a level determined by jitter alone. Taking advantage of variability due to interstellar scintillation, \citet{lcc+16} used this technique to measure the magnitude of jitter noise for a number of millisecond pulsars timed by the NANOGrav collaboration. A later study \citep{lma+19} extended this analysis to measure the frequency-dependence of jitter noise in the NANOGrav 12.5-year data set. More recently, \citet{pbs+21} measured jitter in pulsars observed by MeerKAT.

\begin{figure}
\centering
\includegraphics[width=\textwidth]{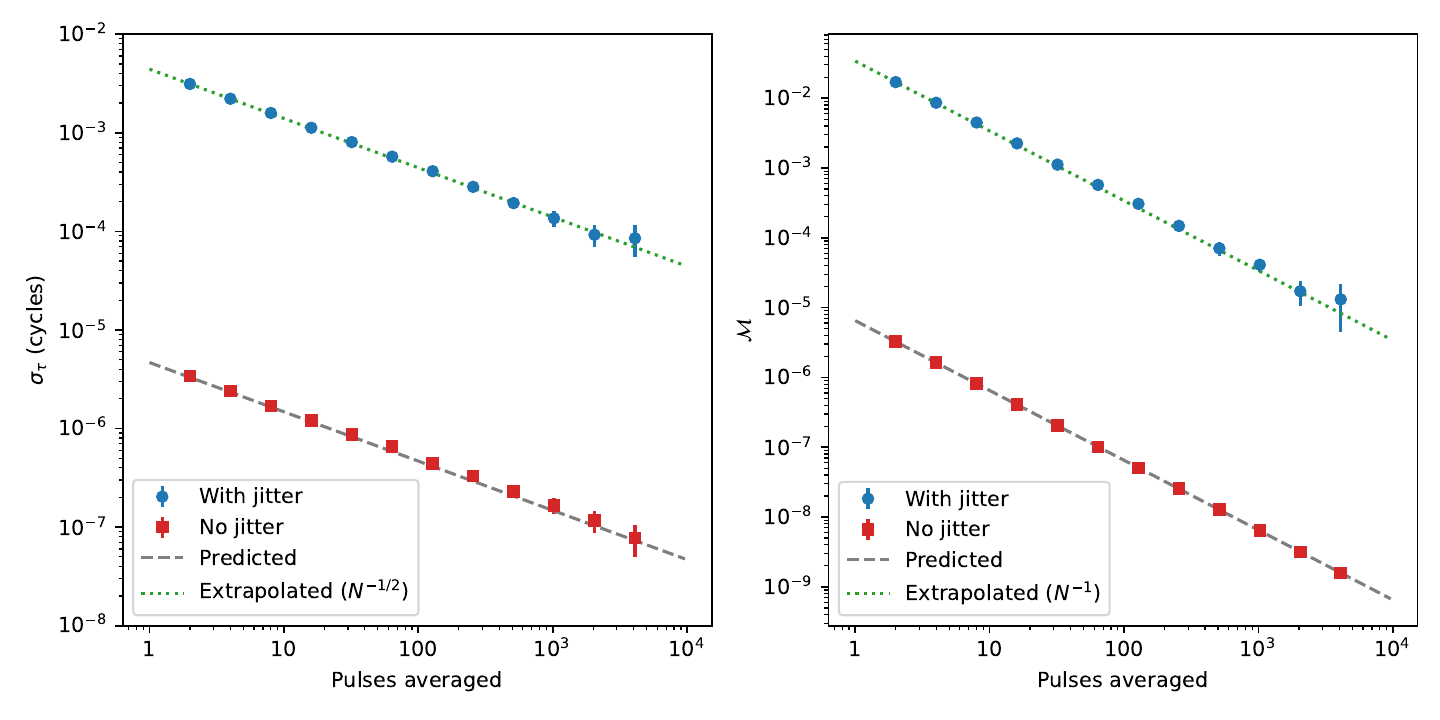}
\caption{The standard deviation of TOA residuals (left) and profile-template mismatch, $\mc{M}$, (right) for average profiles constructed from various numbers of simulated pulses with amplitude and phase variations. In each of two cases, a reference case without jitter and a test case with jitter, 16384 pulses were simulated and grouped together to create profiles with a specific number of pulses each. For $N=2^k$, $k=1,\ldots,12$, the pulses were divided into groups of $N$, and a profile was constructed by averaging the pulses in each group. A TOA was then generated for each profile. The standard deviation of the resulting TOAs is shown in the left panel above, and the average profile-template mismatch is shown in the right panel. In the jitter-free reference case, the results can be predicted using equations~(\ref{eqn:err-from-weff}) and~(\ref{eqn:mismatch-estimate}), respectively; these predictions are shown as dotted gray lines. The dotted green line in each panel is an extrapolation based on the value for $N=2$ and the expected scaling law ($N^{-1/2}$ for the TOA residuals or $N^{-1}$ for the mismatch). The fact that the points from the test case fall along this line, in both cases, demonstrates that the scaling laws hold for this simulated data.}
\label{fig:simulated-sigma-toa-mismatch}
\end{figure}

If the shapes of successive pulses are uncorrelated, the magnitude of jitter noise in an average profile should be proportional to $N^{-1/2}$, where $N$ is the number of pulses averaged \citep[cf.][]{sc12,sod+14}. Error from radiometer noise alone depends on $N$ in the same way. As a result, the relative contribution of jitter noise to TOA errors is determined by the single-pulse signal-to-noise ratio alone -- averaging a larger number of pulses does not change it. Figure~\ref{fig:simulated-sigma-toa-mismatch} demonstrates this using simulated single-pulse data.

Figure~\ref{fig:simulated-sigma-toa-mismatch} also shows the ``mismatch'' between the template and average profiles, defined in terms of the
correlation coefficient $\rho$, for several values of $N$.  The mismatch, $\mc{M}$, between a template, $u(\phi)$, and a profile, $p(\phi)$, is given by $\mc{M}=1-\rho$, where $\rho$
is the correlation coefficient between the (shifted) template and the profile, i.e.,
\begin{equation}
\rho = \frac{\int_0^1 p(\phi) \mkern1mu u(\phi-\hat\tau)\dd\phi}{\sqrt{\int_0^1 p(\phi)^2 \dd\phi \int_0^1 u(\phi)^2 \dd\phi}},
\end{equation}
and $\hat\tau$ is the matched-filtering TOA estimate. Defining the estimated residual, $\hat{r}(\phi)$, as in equation~(\ref{eqn:estimated-residual}), we have
\begin{equation}\label{eqn:mismatch-estimate}
\mc{M}\approx\frac{\int_0^1 \hat{r}(\phi)^2\dd\phi}{2\int_0^1 u(\phi)^2\dd\phi}.
\end{equation}
Since $\hat{r}(\phi)$ is proportional to $N^{-1/2}$, the mismatch is proportional to $N^{-1}$. Comparing the TOA errors and mismatch values for average profiles created from different numbers of single pulses to expectations for profiles differing from the average only by additive white noise, as is done in Figure~\ref{fig:vela-n-pulses}, demonstrates the relative size of the contribution from single-pulse stochasticity, as well as the fact that it depends on $N$ in the same way as the contribution from additive white noise.

\subsection{Profile Autocorrelation Functions}

\begin{figure}
\centering
\includegraphics[width=0.7\textwidth]{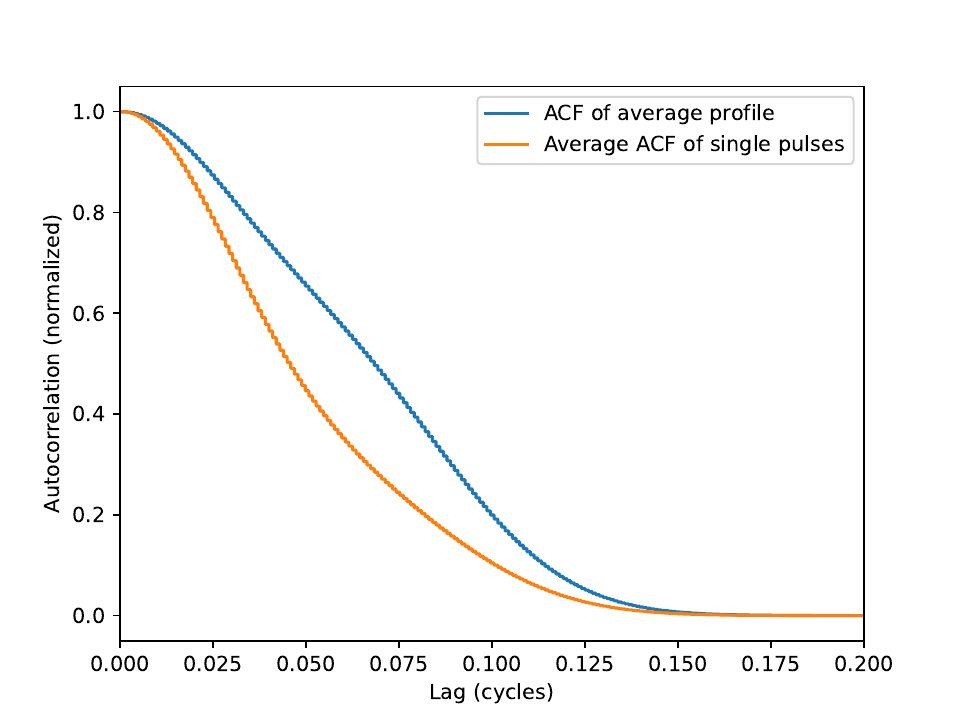}
\caption{Comparison of single-pulse and average-profile autocorrelation functions for simulated pulses with two components of equal amplitude and width $w=0.0425$ cycles, separated by $\delta=0.06$ cycles. Each component is subject to phase jitter with jitter parameter $f=0.3$, and amplitude modulation with modulation index $m=1.0$. Estimating the overall jitter parameter using equation~(\ref{eqn:fhat-acf}) gives $\hat{f}=0.55$.}
\label{fig:simulated-autocorr-comparison}
\end{figure}

Another way to characterize single-pulse stochasticity is in terms of the phase autocorrelation function (ACF), $R(\tau)$, of a profile, $p(\phi)$, defined as
\begin{equation}
R(\tau) = \frac1{\bracket{p^2}}\int_0^1 p(\phi)p(\phi-\tau)\dd{t}.
\end{equation}
Here, as in Section~\ref{sec:toa-estimation}, we define $\bracket{p^2}=\int_0^1 p(\phi)^2 \dd\phi$. The ACF is particularly useful as a measure of pulse shape because it does not depend on how the profile is aligned in phase. Its width, which is a measure of the profile width, tends to be larger for average profiles than for single pulses. This is a consequence of single-pulse stochasticity: the average of many individual narrow pulses as slightly different phases is a broader pulse. This is illustrated using simulated data in Figure~\ref{fig:simulated-autocorr-comparison}.

For a single-component pulse with RMS width $w$ which is subject to phase jitter with standard deviation $fw$, the RMS width of the single-pulse ACF will be $W_{\mr{sp}} = \sqrt{2}w$, while the RMS width of the average-profile ACF will be $W_{\mr{avg}} = \sqrt{2(1+f^2)}w$. As a result, one can estimate the jitter parameter, $f$, as
\begin{equation}\label{eqn:fhat-acf}
\hat{f} = \sqrt{\frac{W_{\mr{avg}}^2}{W_{\mr{sp}}^2}-1},
\end{equation}
where $W_{\mr{sp}}$ and $W_{\mr{avg}}$ are the RMS widths of the single-pulse and average-profile ACFs, respectively.

\subsection{Correlations Between Frequency Channels}

Because pulse-to-pulse shape changes are broadband, the associated TOA estimation errors are strongly correlated between nearby frequency channels. If TOAs are calculated for many frequency channels within the same observation, this characterization can serve as another way to separate TOA estimation errors from single-pulse stochasticity from other sources of noise, and thereby estimate their magnitude. Such estimates can be used to inform pulsar timing models. For example, \citet{5yr-cw} and subsequent NANOGrav analyses \citep[e.g.][]{12yr-nb-timing,15yr-detchar} employ a model of the form
\begin{equation}\label{eqn:efac-equad-ecorr}
    C_{\nu\nu',tt'} = \mc{F}^2\paren{\sigma_r^2 + \mc{Q}^2}\delta_{\nu\nu'}\delta_{tt'} + \mc{J}^2\delta_{tt'}
\end{equation}
for the covariance between TOA residuals at times $t$ and $t'$ and frequencies $\nu$ and $\nu'$. Here, $\sigma_r$ is the estimated TOA uncertainty based on radiometer noise alone (equation~\ref{eqn:sigma-tau}), and the noise parameters $\mc{F}$, $\mc{Q}$, and $\mc{J}$, representing, respectively, an uncertainty scaling factor (EFAC), additional white noise added in quadrature (EQUAD), and a white noise component correlated in frequency but uncorrelated in time (ECORR), are inferred alongside the timing model parameters. The inclusion of the correlated-noise parameter, $\mc{J}$, allows for TOA estimation errors of the sort described here to be taken into account. However, estimates of $\mc{J}$, which are based on TOAs alone, represent a limited view, and are best understood in combination with one or more of the other metrics presented here.

\subsection{Principal Component Analysis}

\begin{figure}
\centering
\gridline{
    \fig{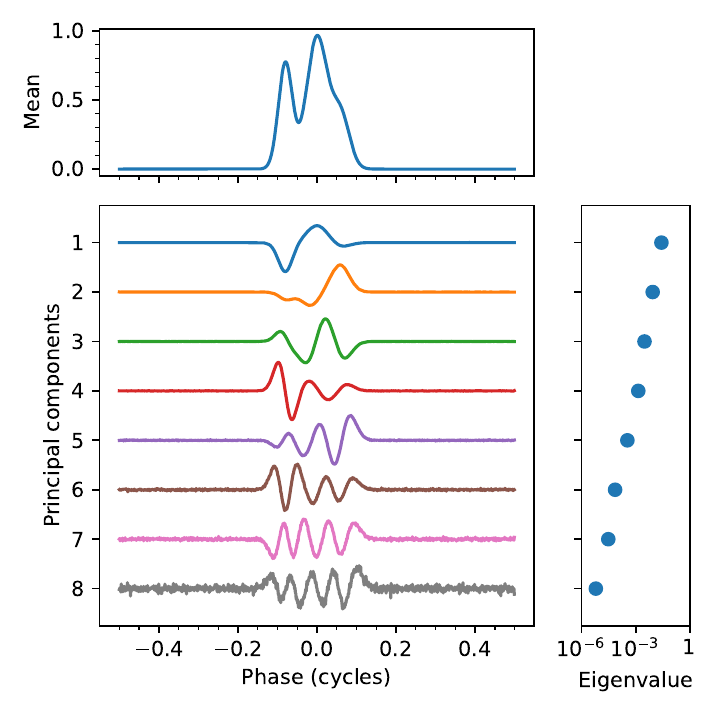}{0.5\textwidth}{(a) A three-component profile with amplitude and phase variations.}
    \fig{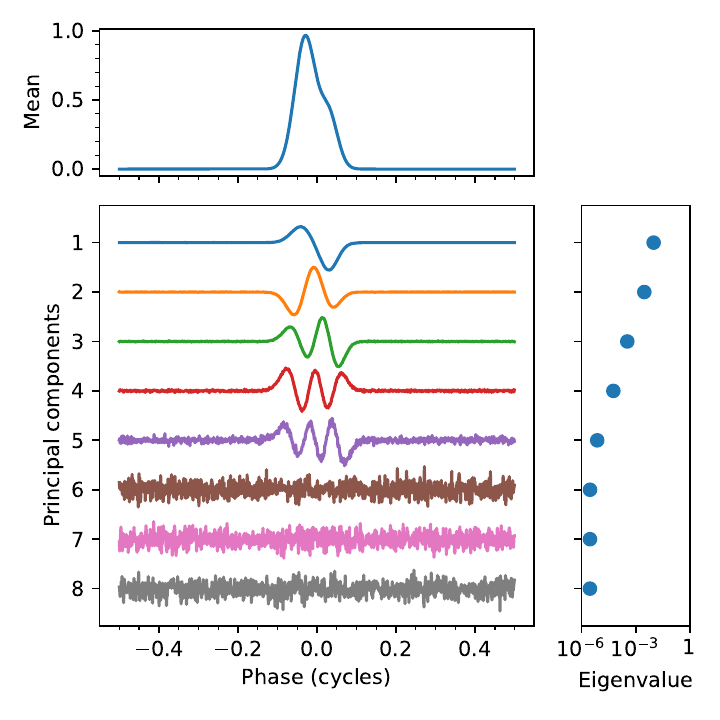}{0.5\textwidth}{(b) A two-component profile with amplitude and phase variations.}
}
\gridline{
    \fig{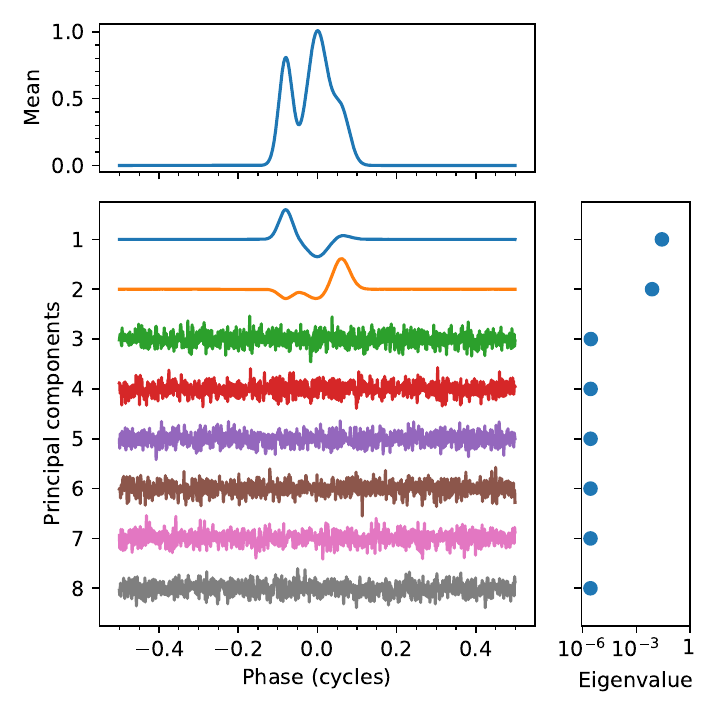}{0.5\textwidth}{(c) A three-component profile with only amplitude variations.}
    \fig{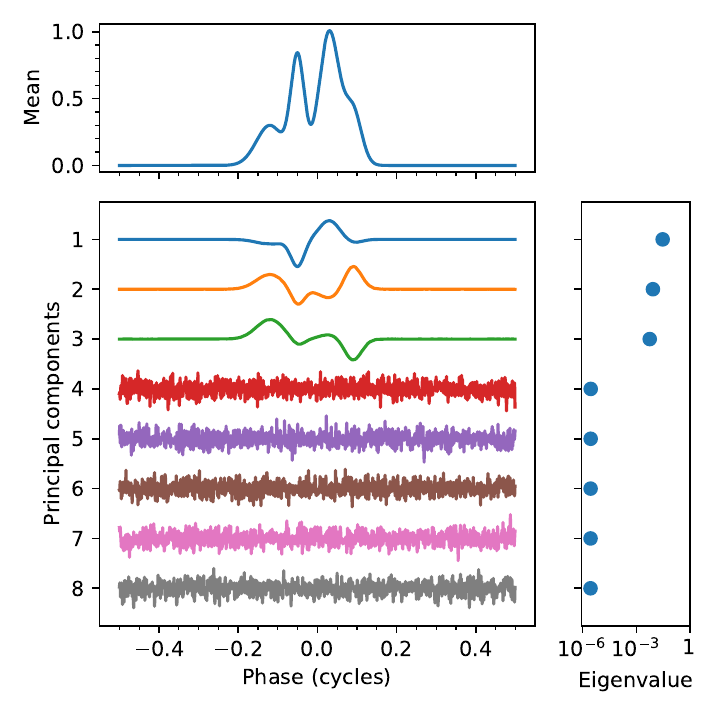}{0.5\textwidth}{(d) A four-component profile with only amplitude variations.}
}
\caption{Principal components in four simulated cases. Within each case, the upper panel shows the average profile, the lower left panel shows the principal components, ordered by eigenvalue from greatest to least, and the lower right panel shows the eigenvalues, using a logarithmic scale. In cases (a) and~(b), amplitude and phase variations (with amplitude modulation index $m=1.0$ and phase jitter parameter $f=0.3$) were added to each component of the profile, whereas, in cases (c) and~(d), only amplitude variation (with $m=1.0$) were added. Cases (c) and~(d) demonstrate that, when only amplitude variations are present, the number of significant principal components is equal to $N_c-1$, where $N_c$ is the number of independently varying profile components. When phase variations are present, however, such as in cases (a) and~(b), additional significant principal components are present, and the line between significant and insignificant principal components is less sharp.}
\label{fig:pcs-four-cases}
\end{figure}

\begin{figure}
\centering
\includegraphics[width=\textwidth]{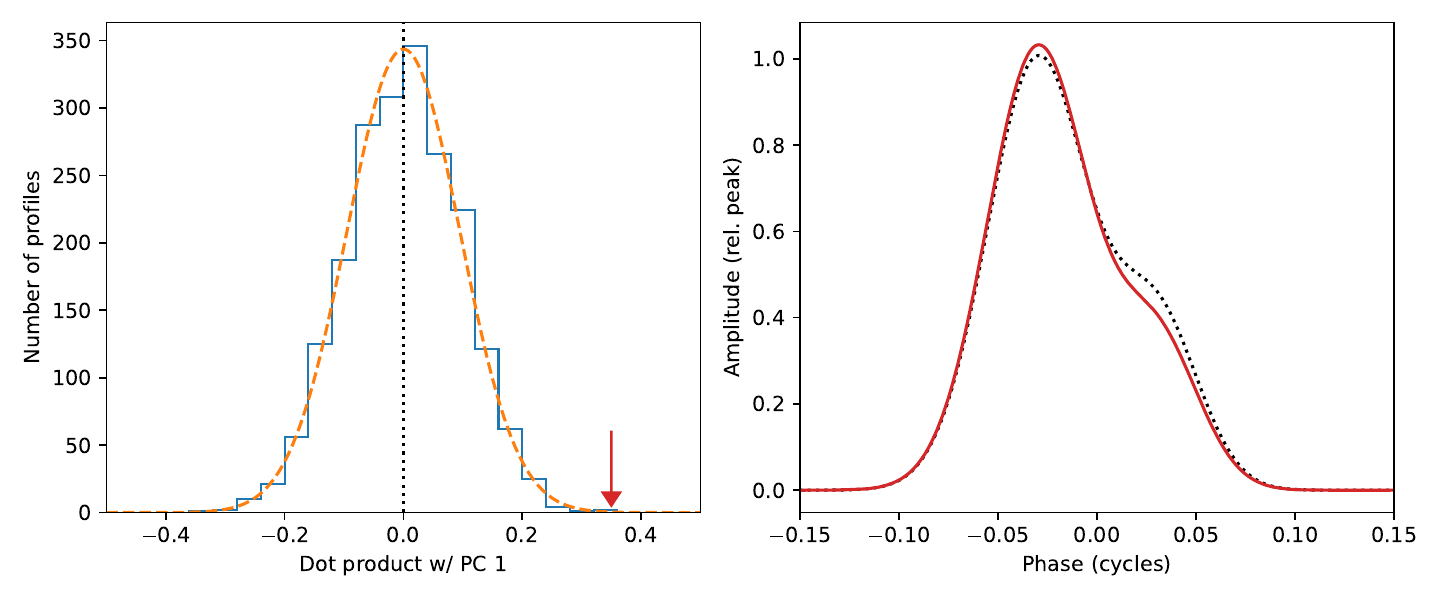}
\caption{The correspondence between shape variations and principal component dot products. The left panel shows a histogram of the dot product $x_1$ with the first principal component for $N_p=2048$ simulated profiles, each constructed from $N=1000$ simulated pulses. The dashed orange line gives the expected Gaussian distribution of dot product values, whose width is the square root of the eigenvalue, $\sigma_1^2$. The right panel shows a profile (red) corresponding to a large value of the dot product, indicated by an arrow in the left panel. The template shape (dotted black line) is included for comparison. The data shown are for a set of simulated profiles with two Gaussian components, each with modulation index $m=1$ and no phase jitter ($f=0$). The leading component, centered at $\phi=-0.03$, has a FWHM of $0.06$ cycles and a mean amplitude of $1.0$, while the trailing component, at $\phi=0.03$, has a FWHM of $0.05$ cycles and a mean amplitude of $0.4$.}
\label{fig:my_label}
\end{figure}

A more direct way of analyzing variations in the shapes of the profiles is to perform principal component analysis (PCA) on the profile residuals. PCA finds the set of orthogonal and statistically independent modes of variation which most effectively captures the variance in the data. Specifically, it results in a set of $k$ principal components, $v_i(\phi)$, and associated variances, $\sigma_i^2$, which allow the profile residuals to be represented as
\begin{equation}
r(\phi) = \sum_{i=1}^{K} x_i v_i(\phi),
\end{equation}
where the coefficients $x_i$ are statistically independent and have variance $\sigma_i^2$:
\begin{equation}
\bracket{x_ix_j}=\sigma_i^2\delta_{ij}
\end{equation}
and the principal components $v_i(\phi)$ are orthonormal:
\begin{equation}
\int_0^1 v_i(\phi) v_j(\phi)\dd\phi = \delta_{ij}.
\end{equation}
The principal components are constructed from the data in a way that maximizes the variance ``explained'' by the first few: the first principal component represents the direction in which the data variance is largest, the second represents the direction, orthogonal to the first, in which the remaining variance is the largest, and so on. The idea of using PCA to characterize jitter was introduced in \citet{demorest-thesis}, and further developed by \citet{ovh+11}, who also considered its use as a correction technique, something we intend to explore further in future work.

One can find the principal components of a set of profile residuals, $r_i(\phi)$, by calculating the singular value decomposition (SVD) of the $N_\phi\times N$ matrix $\mat{R}$ with the residuals as columns, i.e., with entries $R_{ij}=r_j(\phi_i)$, where $\phi_i$ is the phase value corresponding to phase bin $i$. The SVD gives the factorization
\begin{equation}\label{eqn:svd}
\mat{R} = \sqrt{N}\mkern2mu\mat{U}\mat{S}\mat{V}^\T,
\end{equation}
where $N$ is the number of residuals, $\mat{S}=\operatorname{diag}(\sigma_i)$ is a diagonal matrix whose nonzero entries are the square roots of the variances, $\sigma_i^2$, and $\mat{U}$ and $\mat{V}$ are orthogonal matrices, i.e., $\mat{U}^\T\mat{U}=\mat{V}^\T\mat{V}=\mat{I}$. The total number, $K$, of nonzero variances, and thus of principal components, is equal to the rank of the matrix $\mat{R}$, which is almost certainly equal to whichever of $N$ or $N_\phi$ is smaller. The columns of $\mat{V}$ are the principal components, $v_i(\phi)$, i.e., its entries are $V_{ij}=v_j(\phi_i)$. Equivalently, one may also compute the eigenvalue decomposition of the covariance matrix of the residuals,
\begin{equation}\label{eqn:covmat}
\mat{C} = \frac1N\mat{R}\mat{R}^\T,
\end{equation}
which, using equation~\ref{eqn:svd}, is given by
\begin{equation}
\mat{C} = \mat{V}\mat{S}^2\mat{V}^\T.
\end{equation}
This is a somewhat less numerically stable computation, but may be faster if the number of residuals, $N$, is significantly greater than the number of phase bins, $N_\phi$. It also reveals an important property of the principal components. If the pulses from a pulsar are stationary at fixed pulse phase (which they are generally assumed to be, in the absence of mode changes), the covariance matrix in equation~(\ref{eqn:covmat}) is a stable property of the pulsar, and so the principal components are as well.

In simple cases, the shape and number of the significant principal components is directly related to the type of variation present in the data. Consider the case of simulated pulses with independent amplitude variations in each of several components, but no other shape variations. In this case, the significant principal components, together with the average pulse shape, form a basis for the space spanned by the components of the profile. Since the dimension of this space is equal to the number of profile components, the number of significant principal components is one less than the number of profile components.

For a single-component pulse with only phase variations, the profile is given by
\begin{equation}
p(\phi) = \frac{a}{N}\sum_{i=1}^{N} c(\phi-\psi_i),
\end{equation}
which is a special case of equation~(\ref{eqn:profile-components}). Expanding this in a power series in the phase variations, $\psi_i$, shows that the residual is given by
\begin{equation}\label{eqn:resid-from-phase-var}
r(\phi) = a\mkern1mu \bar{\psi}\mkern1mu c'(\phi) + \mc{O}(\psi_i^2),
\end{equation}
where $\bar\psi$ is the mean of the phase variations. The dominant principal component in this case, then, should be proportional to $c'(\phi)$. This component is most dominant when $\bar\psi$ is much less than the width of the pulse; i.e., $f_j\ll 1$. For larger values of $f_j$, the higher-order terms in equation~(\ref{eqn:resid-from-phase-var}) become increasingly important, and contribute additional significant principal components related to higher derivatives of the component shape $c(\phi)$. These tend to have larger numbers of zero crossings as they become less significant, which is primarily results from the fact that each new principal component must be orthogonal to all of the previous ones.

Principal components for several simulated cases are shown in Figure~\ref{fig:pcs-four-cases}, demonstrating the relationships described above: in the absence of phase variations, the number of significant principal components is one less than the number of independently varying components in the profile, and in the presence of phase variations, additional significant principal components are present, with the less-significant principal components having a larger number of zero crossings.

PCA is a very sensitive probe of subtle variations in large data sets. This means that isolating the characteristic principal components for a given pulsar requires a very clean data set, from which RFI has been thoroughly excised. It also means that PCA can be used as a tool for identifying low-level RFI in places where it might not otherwise be obvious, an idea which will be explored further below. An important tool in this sort of analysis is the set of principal component dot products. Given a set of residuals, $r_i(\phi)$, the principal component dot products are the numbers
\begin{equation}
x_{ij} = \int_0^1 r_i(\phi)\mkern2mu v_j(\phi)\dd\phi.
\end{equation}
Since the principal components form an orthonormal basis for the space spanned by the residuals, these can be used to expand the residuals:
\begin{equation}
r_i(\phi) = \sum_{j=1}^k x_{ij} v_j(\phi).
\end{equation}
In other words, $x_{ij}$ quantifies the extent to which the principal component $v_j(\phi)$ contributes to the residual $r_i(\phi)$. By examining the dot products $x_{ij}$ for a fixed principal component $v_j(\phi)$, one can find the profiles it contributes to the most, which is particularly useful if $v_j(\phi)$ represents an identifiable feature in the data.

\subsection{Application to the Vela Pulsar}

\begin{figure}
\centering
\includegraphics[width=\textwidth]{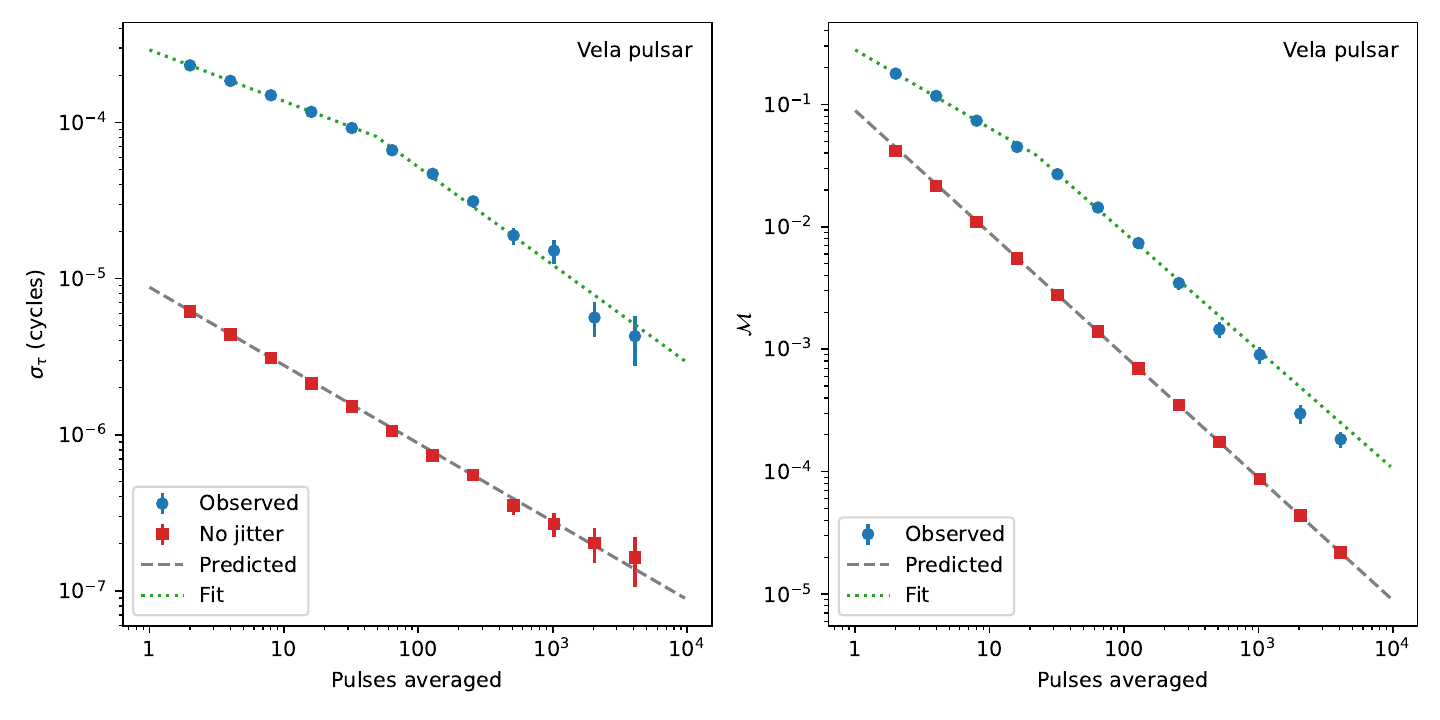}
\caption{The standard deviation of TOA residuals (left) and profile-template mismatch (right) for average profiles constructed from various numbers of single pulses from the Vela pulsar. This can be compared with Figure~\ref{fig:simulated-sigma-toa-mismatch}, which shows the results of the same analysis for a simulated data set. The results seen above are derived from $2^{14}=16,384$ consecutive single pulses, representing approximately 23 minutes of observing time. For each power-of-two value of $N$ between $2$ and $4096$, profiles were constructed from groups of $N$ consecutive pulses and the corresponding TOAs and mismatch values computed. Unlike in the simulated case, the results here deviate noticeably from the expected power laws for averages consisting of only a small number of pulses. A broken power law was fit to the data in each panel; the best-fit model is shown as a green dotted line. The power-law indices for $\sigma_\tau$ were $-0.328\pm0.007$ and $-0.63\pm0.03$ for low and high pulse number, respectively, with the break occurring at $N=48\pm7$. For the mismatch $\mc{M}$, the power-law indices were $-0.638\pm0.009$ and $-0.96\pm0.09$, respectively, with the break occurring at $N=22\pm6$.
}
\label{fig:vela-n-pulses}
\end{figure}

\begin{figure}
\centering
\includegraphics[width=0.7\textwidth]{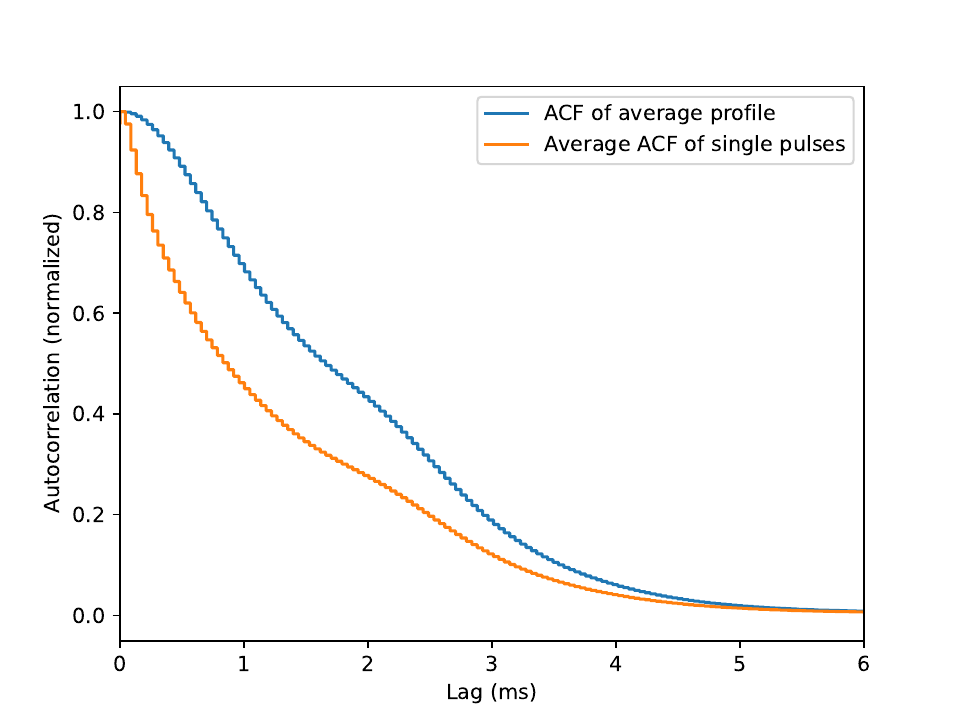}
\caption{Comparison of single-pulse and average-profile autocorrelation functions derived from 18,254 single pulses from the Vela pulsar. As in Figure~\ref{fig:simulated-autocorr-comparison}, the ACF of the average profile is significantly broader. Estimating the overall jitter parameter using equation~(\ref{eqn:fhat-acf}) gives $\hat{f}=0.28$.}
\label{fig:vela-autocorr-comparison}
\end{figure}

\begin{figure}
\centering
\includegraphics[width=\textwidth]{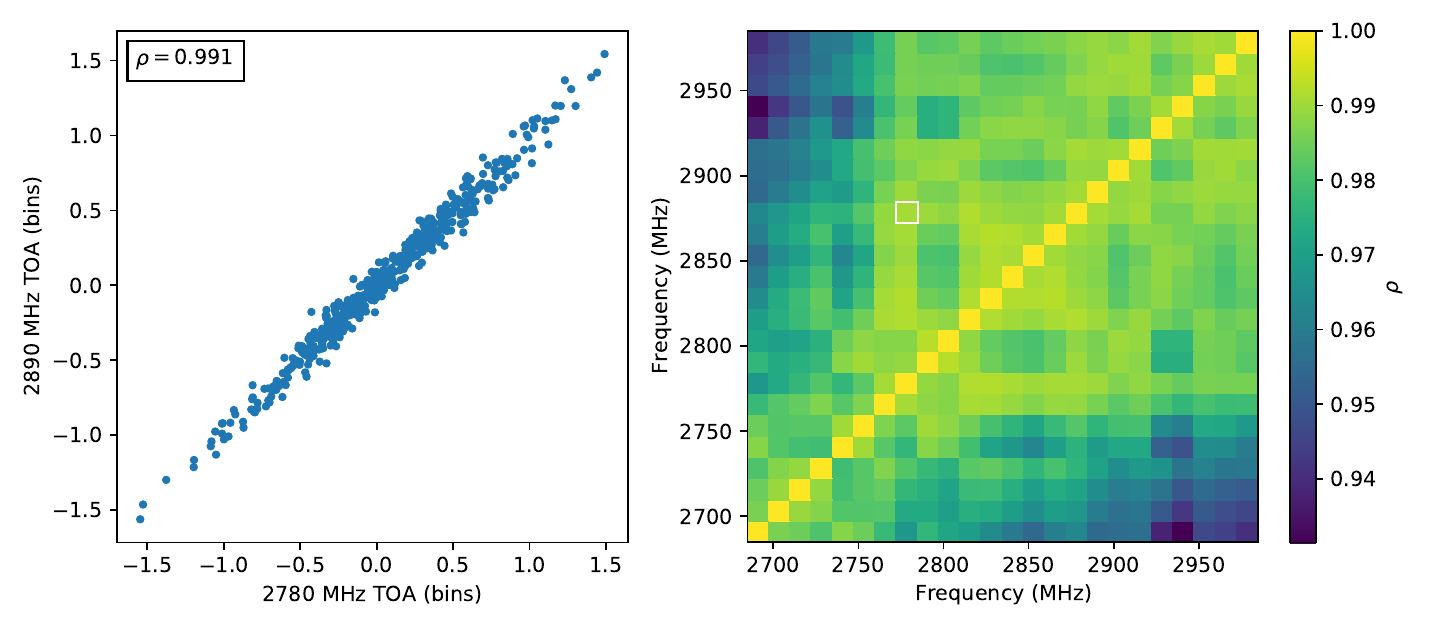}
\caption{Correlations between TOAs in nearby frequency channels for the Vela pulsar. Here the observing band was divided into 24 channels, and a correlation coefficient was calculated between the TOA estimates for each pair of channels. The TOAs for a single pair of channels are shown in the left panel. TOAs are measured in phase bins, where one bin corresponds to approximately \SI{43.6}{\micro s}. The right panel shows correlations between all channel pairs, with a small white square identifying the pair used in the left panel. Notably, even the weakest correlation here is still very strong, at 0.93, indicating that the single pulses decorrelate only marginally  over \SI{300}{MHz}.}
\label{fig:vela-freq-correlations}
\end{figure}

\begin{figure}
\centering
\includegraphics[width=0.625\textwidth]{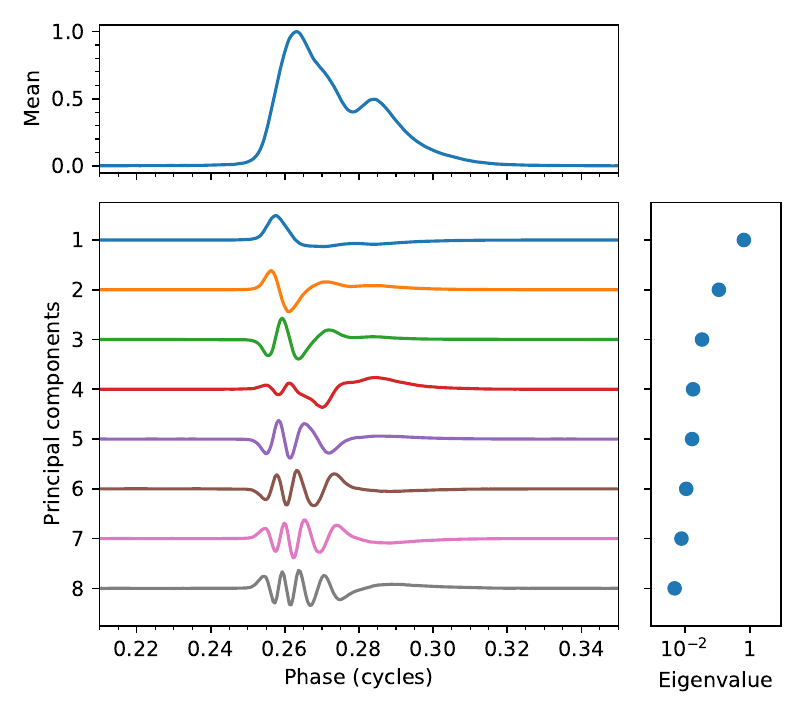}
\caption{Principal components derived from single pulses from a May 4, 2016 observation of the Vela pulsar.}
\label{fig:vela-pcs}
\end{figure}

\begin{figure}
\centering
\includegraphics[width=\textwidth]{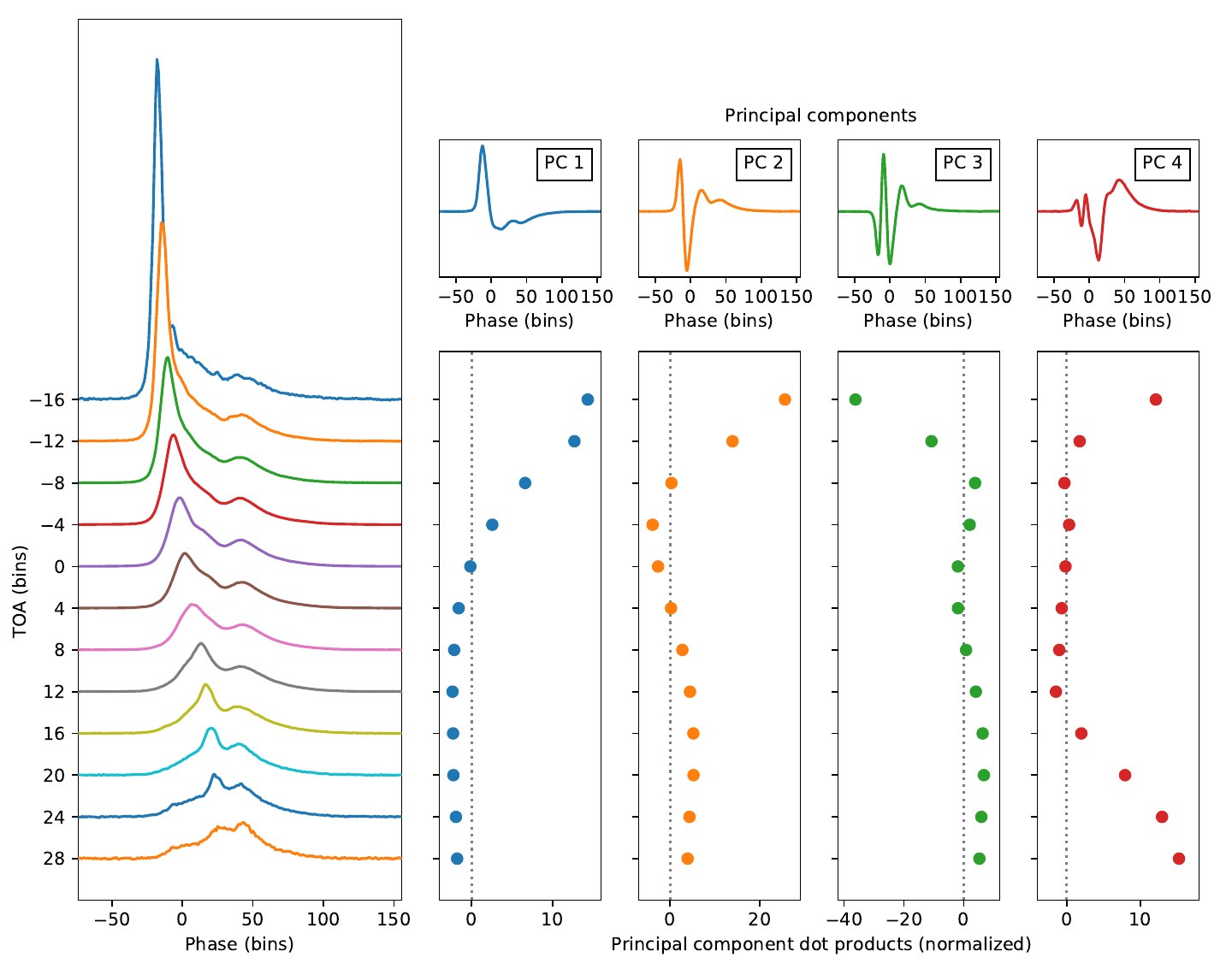}
\caption{An illustration of the relationship between pulse shape and the matched-filtering estimate of the TOA for the Vela pulsar. Data are from MJD 57532 (May 4, 2016). Pulses were divided into groups based on their relative arrival times. The average pulse shape in each group is shown in the left panel above. The correlation between arrival time and intensity identified by \citet{kd83} is clearly visible. The panels to the right show the dot products of the first four principal components (see Figure~\ref{fig:vela-pcs}) with the group average pulses, which have been normalized by dividing them by the square roots of the corresponding variances.}
\label{fig:vela-pulses-by-toa}
\end{figure}

The Vela pulsar, PSR B0833$-$45, is one of the brightest radio pulsars. This makes it a particularly good test case for the study of single-pulse stochasticity, since it is possible to observe single pulses at very high signal-to-noise ratios. Vela is a young pulsar, with a period $P=\SI{89.3}{ms}$ and a spindown rate $\dot{P}=\num{1.25e-13}$, giving it a characteristic age $P/2\dot{P}=\SI{11.3}{kyr}$.
Its period falls between those of millisecond pulsars and those of most canonical pulsars. As in the case of the Crab pulsar, this is attributable to its relative youth. Its significance as a particularly bright and young pulsar has made Vela one of the best-studied of all pulsars, and several authors have previously described the behavior of its single pulses. In particular, \citet{kd83} discovered that there is a strong correlation between the intensity of single pulses from Vela and their arrival times, with brighter pulses arriving earlier. More recently, \citet{kerr15} constructed a basis set of pulse shapes by grouping single pulses from Vela with similar shapes, showing that they could be used to partially correct for the TOA errors associated with shape variation.

Here, we characterize pulse shape changes in the Vela pulsar using a set of single pulses, provided by M. Kerr, which were observed with the Parkes telescope on
MJD 57532 (May 4, 2016). The 28-minute observation was made using the 10 cm receiver on the Parkes telescope at a center frequency of \SI{2820}{MHz}, and processed using the CASPSR backend. Baseband voltages from two orthogonal, linearly polarized receivers were sampled at 800 million samples per second, and recorded in 204 largely contiguous 8-second blocks. We synthesized coherently dedispersed filterbanks in all four Stokes parameters from the baseband voltage recordings using DSPSR \citep{dspsr:2011pasa}, producing spectra for each of the \num{18254} single pulses in 512 frequency channels and 2048 phase bins. The spectra were then filtered to remove the bandpass edges, leaving \SI{300}{MHz} of usable bandwidth between \num{2685} and \SI{2985}{MHz}, and integrated over frequency and polarization to produce a set of frequency-averaged total-intensity profiles. 

The standard deviation of the residuals and profile-template mismatch are plotted as a function of the number of pulses in each average in Figure~\ref{fig:vela-n-pulses}. The results resemble the simulated data shown in Figure~\ref{fig:simulated-sigma-toa-mismatch}, with one notable exception: the expected power laws fail to hold for low values of $N$. Fitting a broken power-law model to the data gives a best-fit power-law index of $-0.328\pm 0.007$ for low pulse number, and $-0.63\pm0.03$ for high pulse number, with the transition between the two taking place at $N=48\pm7$. Similarly, for the profile-template mismatch, $\mc{M}$, fitting a broken power-law model gives a power-law index of $-0.638\pm0.009$ for low $N$, and $-0.96\pm0.09$, with the transition taking place at $N=22\pm6$. The fact that both functions are shallower than expected for low $N$ is likely a result of small but measurable correlations between successive pulses. Similar results were previously obtained by R. Shannon using data from the Vela pulsar that had coarser time resolution (JMC private communication).

The single-pulse and average-profile ACFs for the Vela data are shown in Figure~\ref{fig:vela-autocorr-comparison}. Both ACFs were calculated from the same May 4, 2016 observation, consisting of 18254 single pulses. An ACF was computed for each single pulse, and the results were averaged to produce the single-pulse ACF. The pulses were also combined to produce an average profile, from which the average-profile ACF was computed. 
To compare the single-pulse ACF to the average-profile ACF, 
we correct for the variance contributed by additive white noise,
which makes a contribution to $R(0)$ proportional to its variance. In Figure~\ref{fig:vela-autocorr-comparison}, this contribution has been estimated based on the variance of the noise in the off-pulse region, and subtracted prior to normalization. This allows for a more direct comparison between the two ACFs. A clear difference in width is visible, demonstrating the broadening effect of single-pulse stochasticity. Estimating $f$ from the Vela pulsar data shown in Figure~\ref{fig:vela-autocorr-comparison} using equation~(\ref{eqn:fhat-acf}) gives $f=0.28$. However, as will be seen below, a single-component model is not adequate to describe the pulse-to-pulse variation seen in this data.

As expected for errors caused by single-pulse stochasticity, the TOA estimation errors in the Vela data (estimated using differences between the matched-filtering TOA estimates and the predictions of the timing model) are highly correlated between different frequencies within the band. This is demonstrated in Figure~\ref{fig:vela-freq-correlations}, which shows the correlations between TOAs in different channels. To calculate these, the full \SI{300}{MHz} band was split into 24 channels, each \SI{12.5}{MHz} wide, and 507 average profiles, each consisting of 36 pulses, were constructed in each channel. A TOA was then estimated for each profile using the matched-filtering algorithm described in Section~\ref{sec:toa-estimation}, and a Pearson product-moment correlation coefficient was calculated between each pair of channels. The results show that TOAs at the top and bottom of the band are less correlated than those in adjacent channels, but only slightly: even the weakest correlation shown in Figure~\ref{fig:vela-freq-correlations} is very large, at 0.93. This high degree of correlation supports the use of the single-parameter ECORR model (equation~\ref{eqn:efac-equad-ecorr}) for this observation. Over larger bandwidths, a more significant decorrelation should be expected. In MeerKAT observations of PSR J0437$-$4715, \citet{ksr+24} find that an additional parameter (corresponding to a rank-two covariance matrix) is necessary to model the decorrelation of TOA errors caused by jitter.

Principal components computed from an observation of the Vela pulsar are shown in Figure~\ref{fig:vela-pcs}. They are not altogether dissimilar from the principal components seen in the more complex simulated cases in Figure~\ref{fig:pcs-four-cases}, demonstrating that the simulations are useful for modelling of real pulses.
However, the presence of a large number of significant principal components is consistent with the known variability of
 pulses from the Vela pulsar in both amplitude and phase.

To facilitate comparisons between our results and those of \citet{kd83}, we also broke down the pulses into groups by relative arrival time. The results are shown in Figure~\ref{fig:vela-pulses-by-toa}. To do this, we first calculated a matched-filtering TOA estimate, $\hat\tau$ for of the 18,254 pulses in the observation, and then estimated the topocentric period by performing a linear fit to the $\hat\tau$ values. The pulses were then rotated to align them with respect to this estimated period. Pulses were then assigned groups according to the $\hat\tau$ value after rotation, with each group representing a range of four phase bins, or $1/512$ of the pulse period (\SI{174}{\micro s}). The average profiles for the 12 groups containing at least 16 pulses are shown in the left panel of Figure~\ref{fig:vela-pulses-by-toa}. The remaining 283 pulses, which had $\hat\tau$ values less than $-16$ phase bins, or greater than $+30$ phase bins, were not included in any average (in many of these cases, the pulse was not visible above the noise). The results show clear evidence of the correlation between arrival time and intensity identified by~\citet{kd83}. The panels on the right side of Figure~\ref{fig:vela-pulses-by-toa} show the amplitude of each principal component in each of the average pulses shown at left, demonstrating that these can serve as an alternate description of the systematic variation of the pulse shape as a function of the relative arrival time.

\section{State Switching}\label{sec:state-changes}

As previously mentioned in Section~\ref{sec:causes}, some pulsars appear to alternate between two or more different modes of emission with distinct average pulse shapes. Other pulsars exhibit nulling, appearing to cease emitting entirely for short periods of time. Nulling may be thought of as an extreme form of mode changing, in which one or more of the modes shows little to no emission. Indeed, \citet{elg+05} found that PSR B0826$-$34 showed faint emission, with a different average pulse shape, during its apparent nulls, confirming that the apparent nulling was actually mode changing. Similarly, in the MSP J1909$-$3744, \citet{msb+22} found a weak secondary emission mode corresponding to apparent nulls. Reviews of the nulling and mode-changing phenomena may be found in \citet{rankin86} and \citet{wmj07}, and more detailed studies of nulling in particular are given by \citet{biggs92}, \citet{gajjar17}, and \citet{sm21}.

The state transitions seen in nulling and mode-changing pulsars are not periodic. Instead, they exhibit 
stochastic state durations with particular mean values. 
\citet{cordes13b} showed that in most cases, transitions between states in nulling and mode-changing pulsars, as well as those that show drifting subpulses, may be modeled as a Markov process, although in one case, that of PSR B1931+24, a more complex model involving stochastic resonance appeared to be required.  By and large, these effects are much more prominent in canonical pulsars than in MSPs.  

In general, timing of mode-changing pulsars may be carried out using a separate template for each mode, and the phase alignment between profiles for different modes may be determined using observations which extend across mode transitions. Nulling pulsars may not, of course, be timed during nulls, but in other respects can be timed in the same way as other pulsars.

PCA responds to mode changing in the same way that it responds to amplitude modulation: several principal components will be produced, which, together with the average pulse shape, span the same space as the profiles in each emission mode. The dot products of the profile residuals with these principal components will show a pattern of abrupt changes at the transitions between modes. Such an analysis could in principle be used to identify mode changing where it is not obvious by eye, such as when the difference between modes is small.

Abrupt changes like those seen in PSR J1643$-$1224 \citep{slk+16} and PSR J1713+0747 \citep{xhb+21, lam21, jcc+24}, which may be accompanied by a decay toward a final stable state, can be quantified using PCA in a similar way. Any gradual time evolution of the profile, while it might not be confined to a single principal component, will almost certainly be well described using a small number of principal components, as long as the profiles remain continuous functions of pulse phase.

\section{Instrumental Effects and RFI}\label{sec:rfi}

The shape changes described thus far have been astrophysical in origin, but instrumental effects and RFI can also give rise to apparent shape changes. Such effects only get in the way of making astropysical measurements, but it is for exactly this reason that they must be identified where they occur and taken into account. Most often, this is done by discarding the corrupted data entirely, but it is also possible in some cases to account for the effect of the RFI and recover some useful information from the data. Some of the techniques described above can be useful both for identifying data which has been corrupted by RFI and for understanding the extent to which low-level RFI influences timing.

As mentioned previously, PCA is a sensitive probe of low-level variation in a dataset. As such, it can be used to quickly identify parts of a large dataset which are affected by RFI. Significant RFI will often appear in the first few principal components, even if it affects only a single profile in the entire dataset. The affected profile or profiles can then be determined by examining the dot products of profiles with the component in which the RFI appears. As a method for identifying RFI, this can be significantly faster than manual examination of the profiles, and lends itself well to automation, which is important for large datasets such as those assembled by PTAs.


One particularly common form of interference is baseline ripple -- a sinusoidal signal usually most visible in the off-pulse region of a profile, and typically caused by interference related to AC transmission lines, occuring at harmonics of either 50 or 60 Hz, depending on the frequency used for power distribution in the region in which the telescope is located. Because of its predictable shape, it is possible to quantify the size of TOA errors produced by baseline ripple~\citep[cf.][]{jmr+21}. Using equation~(\ref{eqn:projection}), we can show that the profile residual produced by baseline ripple has the form
\begin{equation}
r(\phi) = \Re\sbrack{z\exp\of{2\pi\ii fP\phi}},
\end{equation}
where $z=x+\ii y$ is the ripple amplitude, $P$ is the pulse period and $f$ is the ripple frequency. Substituting this into equation~(\ref{eqn:projection}) gives
\begin{equation}
\delta\tau\approx -\frac{\Re\sbrack{z\int_0^1 u'(\phi-\tau)\exp\of{2\pi\ii fP\phi}\dd\phi}}{\int_0^1 u'(\phi)^2\dd\phi}.
\end{equation}

\section{Summary and Conclusions}

In this paper, we have quantified the ways in which pulse shape changes lead to errors in the estimation of TOAs from pulsars. Shape changes can occur for a number of reasons. All pulsars show stochastic shape changes in the short term, between individual pulses, and others exhibit nulling or mode-changing behavior, or abrupt shape changes whose effect decays over a period of months. Effects related to dispersion and scattering by the ISM may also cause effective shape changes, as might RFI. Regardless of their origin, differences between observed profiles and the template used for TOA determination lead to TOA estimation errors whose magnitude can be predicted using equation~(\ref{eqn:delta-tauhat}), as long as the difference is small and of known form.

We have also characterized the phenomenology of single-pulse shape variations, both by examining the TOA estimation errors produced by a generative model that includes the effects of component amplitude and shape variations, and by applying various assessment tools to both simulated data and observations of the Vela pulsar. A general formula for the magnitude of TOA estimation errors created by component amplitude and phase variations (equation~\ref{eqn:tauhat-variance}) is derived. We find that, while amplitude variations alone cannot cause TOA estimation errors in a single-component pulse, they can do so if there are multiple components that vary independently in amplitude, and that these effects are most pronounced when two or more components overlap substantially. If there are multiple components that are well-separated in phase, the associated TOA estimation errors combine in a manner weighted by the squared amplitude, and inversely by the width, of each component (equation~\ref{eqn:independent-components}). If the components are identical, the estimation errors decrease in proportion to the inverse square root of the number of components.

These results will have significant consequences for pulsar timing experiments, particularly pulsar timing array searches for gravitational waves. The MSPs timed by PTAs exhibit a diverse set of profile shapes (see Figure~\ref{fig:msp-profiles}), and these are known to have significant consequences for timing as a result of equation~(\ref{eqn:err-from-weff}), with sharp, narrow profiles like that of PSR J1909$-$3744 (seen in the left panel of Figure~\ref{fig:msp-profiles}) yielding the best results. The ability to further characterize the timing performance of pulsars by examining the degree to which single-pulse stochasticity affects TOA estimates could be used to help determine whether a pulsar is suitable for inclusion in a PTA, or to determine whether noise parameter estimates are reasonable. Further, the ability to quantify the extent to which differences from the template shape influence TOA estimates could be used to determine when data affected by shape changes represents a problem for timing. 

\bigskip
\noindent As members of the NANOGrav Physics Frontiers Center, the authors have received support from the National Science Foundation (NSF) under award numbers 1430284 and 2020265. This project made use of the Parkes 64-meter ``Murriyang'' radio telescope. Murriyang is part of the Australia Telescope National Facility (grid.421683.a) which is funded by the Australian Government for operation as a National Facility managed by CSIRO. We acknowledge the Wiradjuri people as the traditional owners of the Observatory site. We would particularly like to thank M. Kerr for providing baseband voltage recordings of the Vela pulsar, PSR B0833$-$45, taken as part of the P816 observing project, as well as providing feedback on a draft version of this document.

\software{Numpy~\citep{numpy}, Matplotlib~\citep{matplotlib}, Astropy~\citep{astropy:2013,astropy:2018,astropy:2022}, PSRCHIVE~\citep{psrchive:2004pasa,psrchive:2011ascl}, DSPSR~\citep{dspsr:2010ascl,dspsr:2011pasa}}

\bibliography{characterization-paper}

\end{document}